\documentclass[12pt]{article}
\usepackage{amsmath,amsthm}
\usepackage{bm,amssymb,mathtools}
\usepackage[ruled,vlined]{algorithm2e}
\usepackage{graphicx,psfrag,epsf}
\usepackage{enumerate}
\usepackage[round,authoryear]{natbib}
\usepackage{url} % not crucial - just used below for the URL 
\usepackage[colorlinks,citecolor=blue,urlcolor=blue,linkcolor = blue,backref=page]{hyperref}

\newtheorem{lemma}{Lemma}
\newtheorem{proposition}{Proposition}	

\newtheorem{remark}{Remark}

\newtheorem{assumption}{Assumption}

\newcommand{\pto}{\stackrel{P}{\to}}
\renewcommand{\P}{\mathbb{P}}

\newcommand{\R}{\mathbb{R}}

\newcommand{\blind}{0}

\begin{document}

\def\spacingset#1{\renewcommand{\baselinestretch}%
{#1}\small\normalsize} \spacingset{1}

%%%%%%%%%%%%%%%%%%%%%%%%%%%%%%%%%%%%%%%%%%%%%%%%%%%%%%%%%%%%%%%%%%%%%%%%%%%%%%

\if0\blind
{
  \title{\bf Bootstrapped Edge Count Tests for Nonparametric
  	Two-Sample Inference Under Heterogeneity}
  \author{Trambak Banerjee%\thanks{
    %The authors gratefully acknowledge \textit{please remember to list all relevant funding sources in the unblinded version}}
\hspace{.2cm}\\
Analytics, Information and Operations Management, University of Kansas\\
Bhaswar B. Bhattacharya\hspace{.2cm}\\
Department of Statistics and Data Science, University of Pennsylvania \\
Gourab Mukherjee\hspace{.2cm}\\
Department of Data Sciences and Operations, University of Southern California}
  \maketitle
} \fi

\if1\blind
{
  \bigskip
  \bigskip
  \bigskip
  \begin{center}
    {\LARGE\bf Bootstrapped Edge Count Tests for Nonparametric
    	Two-Sample Inference Under Heterogeneity}
\end{center}
  \medskip
} \fi

\bigskip
\begin{abstract}
Nonparametric two-sample testing is a classical problem in inferential statistics. While modern two-sample tests, such as the edge count test and its variants, can handle multivariate and non-Euclidean data, contemporary gargantuan datasets often exhibit heterogeneity due to the presence of latent subpopulations. Direct application of these tests, without regulating for such heterogeneity, may lead to incorrect statistical decisions. We develop a new nonparametric testing procedure that accurately detects differences between the two samples in the presence of unknown heterogeneity in the data generation process. Our framework handles this latent heterogeneity through a composite null that entertains the possibility that the two samples arise from a mixture distribution with identical component distributions but with possibly different mixing weights. In this regime, we study the asymptotic behavior of weighted edge count test statistic and show that it can be effectively re-calibrated to detect arbitrary deviations from the composite null. For practical implementation we propose a Bootstrapped Weighted Edge Count test which involves a bootstrap-based calibration procedure that can be easily implemented across a wide range of heterogeneous regimes. A comprehensive simulation study and an application to detecting aberrant user behaviors in online games demonstrates the excellent non-asymptotic performance of the proposed test.
\end{abstract}

\noindent%
\vfill

\newpage
\spacingset{1.5} % DON'T change the spacing!
\section{Introduction}
\label{sec:intro}
Nonparametric two-sample testing is a classical problem in inferential statistics. The use of these tests is pervasive across disciplines, such as medicine \citep{farris1999between}, consumer research \citep{folkes1987field}, remote sensing \citep{conradsen2003test} and public policy \citep{rothman2006patient}, where detecting distributional differences between the two samples is germane to the ongoing scientific analysis. Nonparametric two-sample tests like the Kolmogorov-Smirnov test, the Wilcoxon rank-sum test, and the Wald-Wolfowitz runs test are extremely popular tools for analyzing univariate data. Multivariate versions of these tests have their origins in the randomization tests of \cite{chung1958randomization} and in the generalized Kolmogorov-Smirnov test of \cite{bickel1969distribution}. %Multivariate versions of these tests date back to the randomization tests of \cite{chung1958randomization} and to the generalized Kolmogorov-Smirnov test of \cite{bickel1969distribution}.
\cite{friedman1979multivariate} proposed the  first computationally efficient nonparametric two-sample test, the edge count test, for high-dimensional data. Modern versions of the edge count test, such as the weighted and generalized edge count test \citep{chen2017new,chen2018weighted}, can handle multivariate data, and can be applied to any data types as long as an informative similarity measure between the data points can be defined. Besides the edge count tests, several tests based on nearest-neighbor distances \citep{henze1984number,schilling1986multivariate,chen2013ensemble,hall2002permutation,banerjee2020nearest} and matchings  \citep{rosenbaum,mukherjee2022distribution} have been proposed over the years, and used in variety of applications, such as covariate balancing \citep{heller2010sensitivity,heller2010covariatebalance}, change point detection \citep{chen2015graph,shi2017consistent}, gene-set analysis \citep{geneset2012}, microbiome data \citep{callahan2016,holmes2018modern,fukuyama2020}, among others. 

\cite{bhattacharya2019} propose a general framework to study the asymptotic properties of these graph based tests. In addition to the graph based tests, other popular two-sample tests include the energy distance test of \cite{szekely2003statistics} and \cite{szekely2004testing} (see also \citet{baringhaus2004new,aslan2005new,szekely2013energy}), and kernel tests based on the maximum mean discrepancy (see \cite{gretton2007kernel,chwialkowski2015fast,ramdas2015decreasing,ramdas2017wasserstein} and the references therein). %Recently, \cite{deb2021multivariate} and \cite{ghosal2019multivariate} proposed distribution-free two-sample tests based on the concept of multivariate ranks that is defined using the theory of optimal transport. CITE \citep{shi2020distribution} \citep{shi2020universally}. 
Recently, several authors have proposed distribution-free two-sample tests using optimal transport based multivariate ranks (see \cite{deb2021multivariate,ghosal2019multivariate,shi2020distribution1}, and \cite{shi2020universally2} and the references therein). 
%Recently, several authors have proposed distribution-free two-sample tests based on the concept of multivariate ranks. For instance, \cite{deb2021multivariate} and \cite{ghosal2019multivariate} use the theory of optimal transport to define the multivariate ranks whereas \cite{shi2020distribution1} and \cite{shi2020universally2} use the concept of center-outward ranks and signs which are a multivariate generalization of traditional ranks.

While there exists a vast literature on nonparametric two-sample tests, their performance in scenarios where the two samples might contain different heterogeneous structures have not been well-explored before. However, in a host of contemporary data analysis problems (See Ch. 3 of \citet{holmes2018modern}, \citet{rossi2012bayesian}) there is a need to conduct inference in presence of unknown heterogeneity in the data generation process. This is particularly important when there are latent subpopulations in the two populations from which the samples were extracted. Detecting distributional differences across the two samples is challenging in the presence of such heterogeneity because the two samples may differ with respect to the rates with which they arise from the underlying subpopulations. In these settings, direct application of existing two-sample tests, without regulating for the latent heterogeneity in the samples, may lead to incorrect statistical decisions and scientific consequences.
In this article, we develop a new nonparametric testing procedure that can accurately detect if there are differences between the two samples in the presence of latent heterogeneity in the data generation process. We next present two contemporary data examples to motivate the two-sample testing problem under heterogeneity and discuss how existing tests may lead to incorrect decisions in the presence of heterogeneity.

\subsection{Motivating examples for testing under heterogeneity}  
\label{sec:motivation}
\noindent\textbf{Example 1: Detecting shift in consumer sentiment and spending pattern }-- Detecting shifts in consumer sentiment and their spending pattern in response to exogenous economic shocks, such as a pandemic, war or supply chain constraints, is important from the perspective of public policy and businesses operations across different sectors \citep{agarwal2019digital,crouzet2019shocks,bruun2021,bartik2020impact}. Evidence of such changes in spending pattern is used to make policy decisions on the allocation of future resources to tackle the shifting landscape of consumer demand \citep{balisi2021,liguori2020strategies,akpan2022small}.
Data privacy rules, however, often forbid using personally identifiable information, such as individual consumer demographics and spending patterns over time, thus ruling out the possibility of acquiring a rich consumer level panel data %on the purchase patterns
and utilizing sophisticated tools from causal inference to detect changes in spending patterns in response to the event of interest. A two-sample statistical test of hypothesis is often conducted to determine whether the spending pattern of a sample of consumers before the event of interest is significantly different from the spending pattern of an independent sample of consumers during or after the event of interest.
Despite being much less powerful than tests based on comprehensive longitudinal data, the two-sample tests have the advantage of being substantially more privacy-preserving as these non-intrusive tests only need two independent vectors of observations and thus can be implemented across a wide range of contemporary applications.     
However, there are two main challenges for devising a test of hypothesis that can correctly detect such differences in the spending pattern between these two independent samples:
\begin{enumerate}[(1.)]
	\item The two samples may exhibit sample size imbalances which presents a challenging setting for nonparametric two-sample testing on multivariate data \citep{chen2018weighted}.
	\item The consumer base may consist of several heterogeneous subpopulations with respect to their consumption behavior. % \citep{allenby1998heterogeneity}. 
	In business and economic modeling it is now commonplace to model such consumption behavior as a mixture of several distinct consumption patterns \citep{fahey2007conditional,labeeuw2013residential}.%,netzer2008hidden,kumar2011assessing}.
\end{enumerate}
Due to an exogenous shock we can have one of the following three situations regarding the behavior of consumers after the shock: 
\begin{enumerate}[I.]
\item Consumer behavior does not change and maintains the pre-shock levels.%remains the same as before.
\item Consumption behavior changes but there are no new consumption patterns. The change in the consumption distribution is due to changes in the proportions of existing modes of consumption. This is the case where there is switching between the different consumption modes but no new consumption patterns evolve due to the external shock. 
\item Consumption behavior changes and there are new consumption patterns that were previously non-existent.
\end{enumerate}
In this paper, we concentrate on the detection of the third case, Case III, where new consumption patterns emerge post-shock. Accurate and timely identification of Case III is important for it allows researchers to understand if deviant forms of consumption \citep{koskenniemi2021deviant} arise after a critical event, which can then be subsequently studied and analyzed with higher granularity data. Statistically, the main challenge here lies in correctly detecting Case III while not misidentifying Cases I and II as type I error. Two-sample testing procedures in the existing literature (discussed at the beginning of Section \ref{sec:intro}) are designed to distinguish between Cases I and II. We will show that without modifications their direct usage do not produce correct inference for this exercise.  	

\begin{figure}[!t]
\centering
\includegraphics[width=0.9\linewidth]{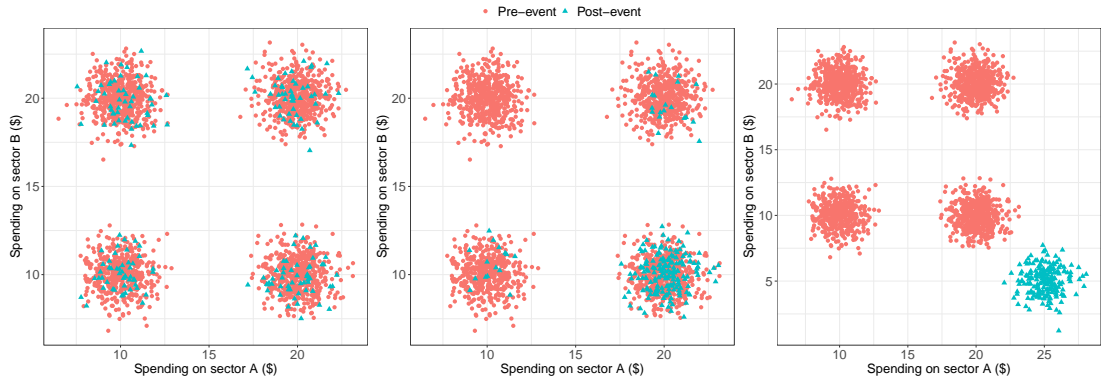}
\caption{\small Testing shift in consumer spending pattern due to external economic shock. Green triangles represent a sample of consumers \texttt{post-event}, while the red dots represent an independent sample of consumers \texttt{pre-event}. The \texttt{pre-event} sample exhibits four distinct subpopulations. \textbf{Left}-- consumption pattern of \texttt{post-event} consumers originate from these four subpopulations with equal probabilities. \textbf{Center}--  majority of the \texttt{post-event} consumers arise from one of the four subpopulations without a shift in the levels of their consumption for that subpopulation. \textbf{Right}-- \texttt{post-event} consumers originate from a new subpopulation that exhibits a change in the levels of consumption after the shock. The simulation scheme for these plots is described in Section \ref{app:figure_1_2_details} of the supplement.}
\label{fig:motfig_1} 
\end{figure}

We further elucidate this problem with the help of a simple simulation example. Consider a bivariate consumption problem with consumption on sector A denoted by $X_1$ and on sector B by $X_2$. We have bivariate observations $(X_1,X_2)$ for two random samples of customers before and after the shock (event).  In Figure \ref{fig:motfig_1}, the red dots represent the sample of consumers \texttt{pre-event}, while the green triangles represent an independent sample of consumers \texttt{post-event}. The three plots in Figure \ref{fig:motfig_1} present the distribution of spending pattern with respect to $(X_1,X_2)$ and reveals that the \texttt{pre-event} sample exhibits four distinct subpopulations. The leftmost panel represents the setting where the consumption pattern of \texttt{post-event} consumers originate from these four subpopulations with equal probabilities. The center panel presents the setting of Case II where a majority of the \texttt{post-event} consumers arise from one of the four subpopulations without a shift in the levels of their consumption for that subpopulation. This represents normal consumption with only the proportional representation of the latent states of normal consumption being changed. The rightmost panel, on the other hand, reveals that the \texttt{post-event} consumers originate from a new subpopulation that exhibits a change in the levels of consumption after the event (Case III).

Distinction between the two settings presented in the center (Case II) and right (Case III) panels of Figure \ref{fig:motfig_1} is critical for policy makers and marketeers. In this paper, we develop a consistent two-sample testing framework that can detect Case III from Cases I and II in the presence of sample size imbalance as well as heterogeneity. We next discuss another motivating example regarding consumption of digital entertainment, which will be further pursued in Section \ref{sec:realdata}. 

%Of course, this goes to higher dimension of vectors;
\noindent\textbf{Example 2: Detecting differences in player behavior in online games }-- Online gaming is an important component of modern recreational and socialization media \citep{banerjeejoint2022}. For monetization of these digital products, managers often have to deliver personalized promotions to users based on their product usage. Additionally, through promotional intervention the portal needs to regulate addiction, violence and other deviant consumption patterns \citep{hull2014longitudinal} than can cause long term societal harms. As it is difficult to manually track every game, the portals use automated decision rules to monitor the game sessions and rely on features extracted at regular intervals of time, such as hourly or half-hourly, to constantly check for deviant consumption within each game session. % based on features extracted over hourly (half-hourly or quarterly hourly) gaming sessions to constantly check for deviant consumption. 
When evidence of preponderance of deviant consumption is available, managerial interventions are made.
\begin{figure}[!h]
\centering
\includegraphics[width=.33\textwidth]{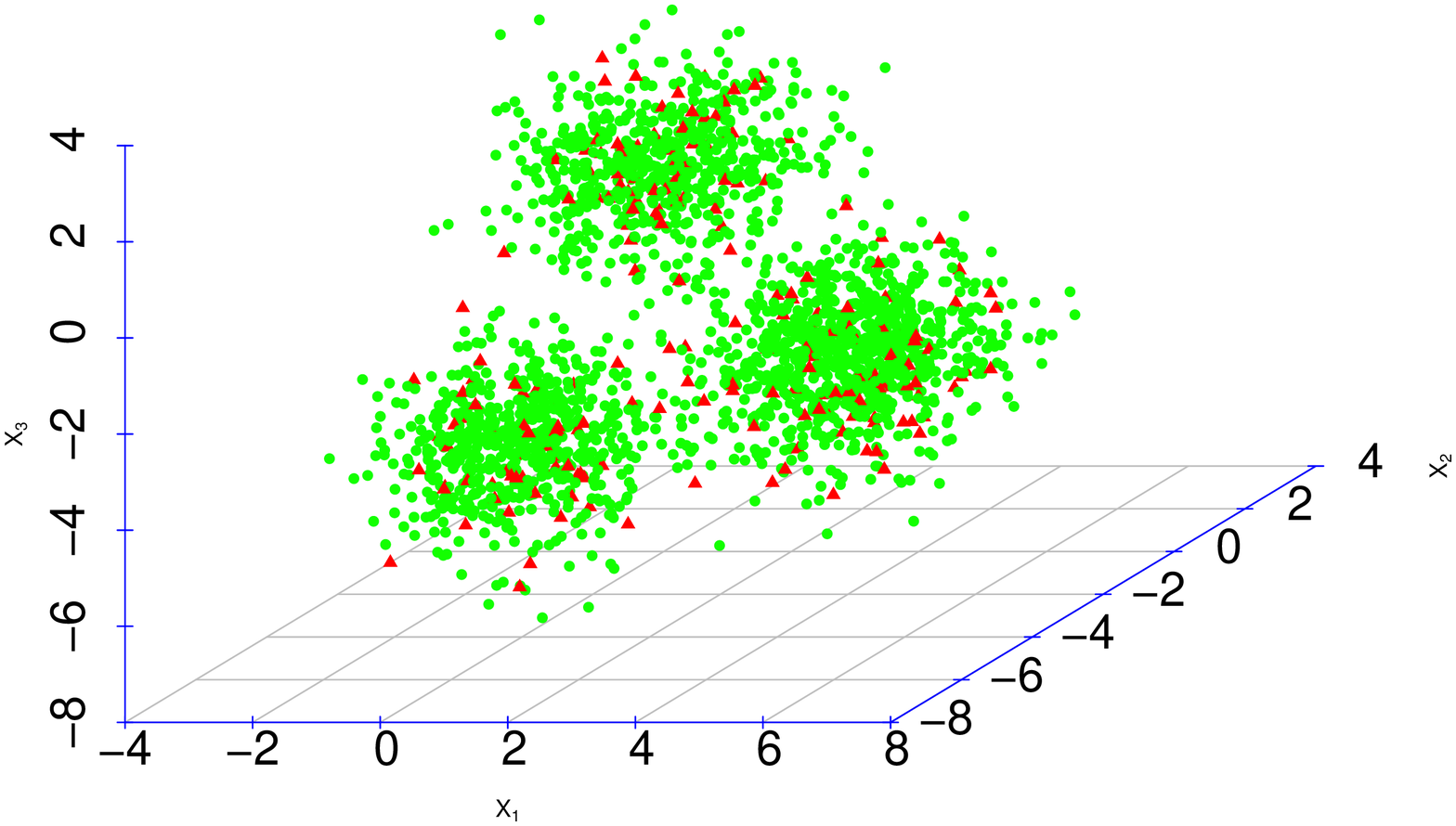}\hfill
\includegraphics[width=.34\textwidth]{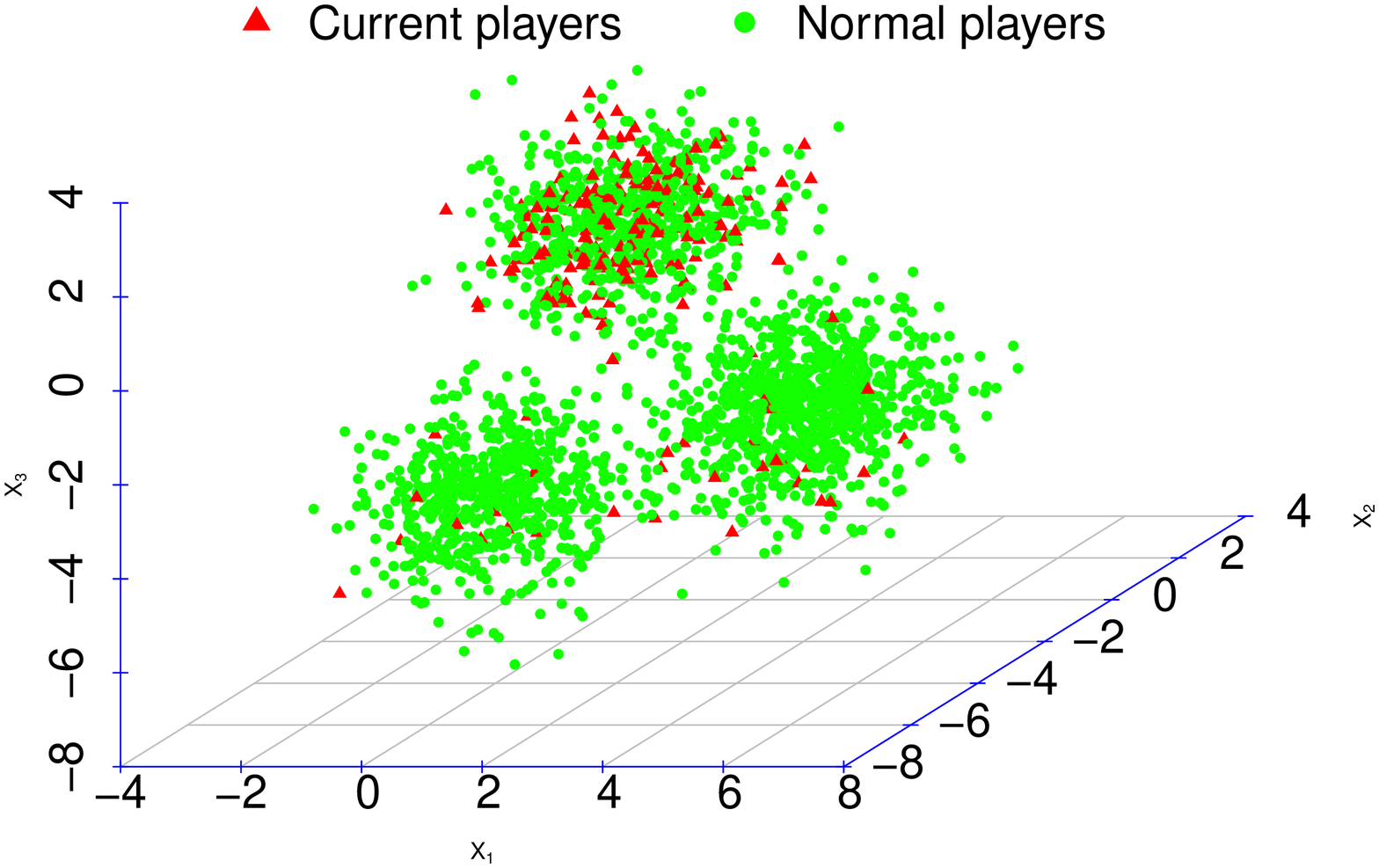}\hfill
\includegraphics[width=.33\textwidth]{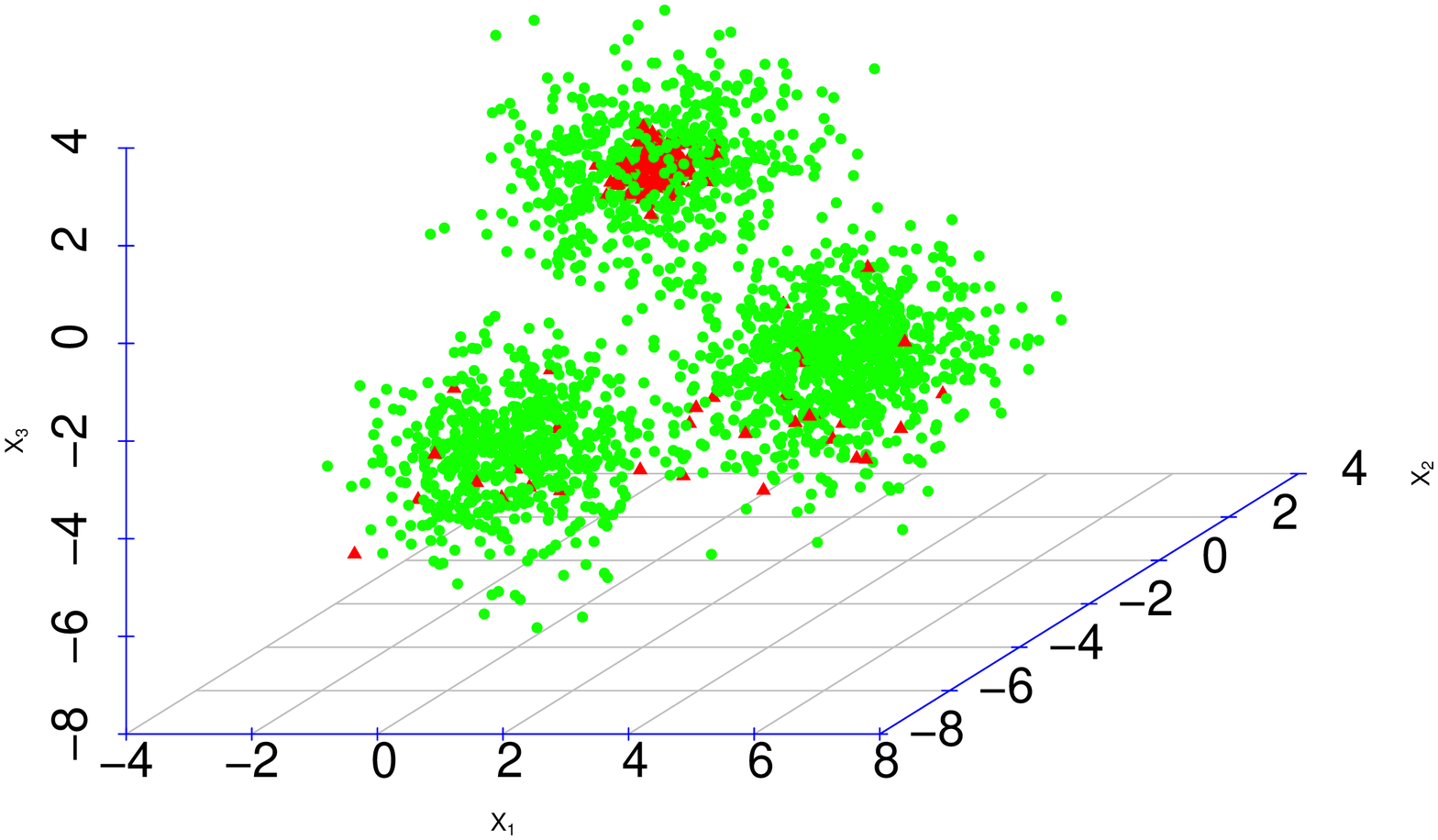}
\caption{\small Differences in playing behavior in online video games. Red triangles represent the sample of current players, while the green dots are the sample of normal players from historical logs. The green sample exhibits three distinct subpopulations. \textbf{Left}-- Case I: {Current sample} originates from three subpopulations present in the green sample with equal probabilities. \textbf{Center}-- Case II: most members in the {current sample} originate from only one of the three subpopulations. \textbf{Right}-- Case III: {current sample} originates from one of the three subpopulations but exhibits a different scale. The simulation scheme for these plots is described in Section \ref{app:figure_1_2_details} of the supplement.}
\label{fig:motfig_2}	
\end{figure}

Statistically, the problem here again reduces to the correct detection of scenarios pertaining to Case III described in Figure \ref{fig:motfig_1} above. To see this, note that based on historical data a multivariate sample of gaming features from players corresponding to normal gaming characteristics is available. The size of such a sample is huge as it is based on historical logs. The goal is to compare the instantaneous gaming features of a subset (gated by region, age, etc) of currently logged-on players with respect to this sample. The sample size of this subset of players is much smaller than the benchmark normal gaming consumption sample and so, we encounter the issue of sample size imbalance. Additionally, %Also, note that 
while comparing these two samples we are interested in detecting if there is a sub-population of players with deviant gaming characteristics which would need regulating promotional incentives related to purchase of gaming artifacts. % and life/time.
This again corresponds to different Case III scenarios described above. Even without any instance of deviant usage, instantaneous gaming characteristics greatly change over time-of-day, for instance early morning players and evening players have varying characteristics. This corresponds to scenarios in Case II. Thus, the goal will be to detect Case III scenarios while allowing the possibility for scenarios from Cases I and II. We illustrate this through a simulation example in Figure~\ref{fig:motfig_2}. 

Consider observing $d=3$ dimensional characteristics, $X_1,X_2$ and $X_3$ of playing behavior. The green dots in Figure~\ref{fig:motfig_2} represent the sample of normal players from historical records. It exhibits three distinct subpopulations of equal sizes. The red triangles represent the sample of players from a current session. The leftmost panel depicts a setting where the current sample originates from these three subpopulations with equal probabilities (an instance of Case I). The center panel presents the setting where a majority of the players in the current sample are from one of the three subpopulations (an instance of Case II), and the rightmost panel reveals an instance of Case III where there is a new sub-population represented by red triangles that was non-existent in normal players. 
Note that, unlike Figure~\ref{fig:motfig_1} here the deviant sub-population differs from one of the three subpopulations of normal players only in scale and not in locations. Detecting this case is more difficult than Figure~\ref{fig:motfig_1} and we present a detailed analysis in Section~\ref{sec:sims}. In Section~\ref{sec:realdata} we develop this example and apply our proposed hypothesis testing method  to detect addictive behaviors in a real-world gaming dataset. 

\subsection{Two-sample testing under heterogeneity and our contributions}
Existing nonparametric two-sample tests cannot distinguish between the scenarios presented in the center and right panels of Figures \ref{fig:motfig_1} and \ref{fig:motfig_2}. For instance, the edge count test \citep{friedman1979multivariate}, the weighted edge count test \citep{chen2018weighted} and the generalized edge count test \citep{chen2017new} % \citep{chen2017new,chen2018weighted,friedman1979multivariate}, the cross-match test \citep{rosenbaum2005exact}, and the energy test \citep{aslan2005new} 
reject the null hypothesis of equality of the two distributions for both Cases II and III in Figures \ref{fig:motfig_1} and \ref{fig:motfig_2} (see tables \ref{tab:motfig_1} and \ref{tab:motfig_2} in Section \ref{app:figure_1_2_details} of the supplement for more details). This is not surprising because these nonparametric two-sample tests are designed to test the simple null hypothesis of equality of the two distributions (Case II vs Case I) and directly using them, without any modification, to detect Case III scenarios from Case II can be misleading. To resolve this conundrum, we develop a hypothesis testing framework based on an appropriately constructed composite null hypothesis (Equations \eqref{eq:setup-1}--\eqref{eq:setup-3}). Under this composite null hypothesis, we study the properties of the Weighted Edge Count (WEC) test statistic of \cite{chen2018weighted} and show that it can be re-calibrated to produce an asymptotically consistent two-sample test. For finite sample applications, we propose a bootstrap-based calibration procedure for the WEC test statistic that allows it to consistently and efficiently detect differences between the two samples under subpopulation level heterogeneity. The ensuing discussion summarizes our key contributions.%Bootstrap based calibration. The proposed Bootstrapped calibrated Weighted Edge Count test statistic, \texttt{BWEC}, is based on the weighted edge count test of \cite{chen2018weighted} that works well under sample size imbalance and also involves a bootstrap based calibration that allows it to consistently and efficiently detect the differences between the two samples under subpopulation level heterogeneity. The ensuing discussion summarizes our key contributions.

%for testing a composite null hypothesis that the two samples arise from a mixture distribution with identical component distributions but with possibly different mixing weights. This is the case for the center panel of Figures \ref{fig:motfig_1} and \ref{fig:motfig_2}. The \texttt{BWEC} test statistic

\begin{itemize}
\item 	%Both the EC test and the WEC tests are universally consistent, that is, they are asymptotically powerful for detecting alternatives, for the classical two-sample with homogeneous populations, but not under heterogeneity.  
We study the asymptotic properties of the WEC statistic for the heterogeneous two-sample problem (Section \ref{sec:2}). Specifically, we show how one can choose a cut-off that renders the WEC statistic asymptotically powerful for detecting any distribution that significantly deviates from the composite null hypothesis (see Equation \eqref{eq:setup-3}), which encompasses the possibility that the two samples arise from the same mixture distribution but with possibly different mixing weights (see Proposition \ref{consistency}). This is in contrast to the TRUH test in \citet{banerjee2020nearest} which can only detect deviations in location under the concerned composite null hypothesis. This is because, unlike the TRUH statistic which compares the nearest neighbor distances between the two-samples, the WEC statistic is based on the number of within sample edges in a similarity graph of the pooled sample. The combinatorial (count-type) nature of the WEC statistic renders it asymptotically distribution-free under the simple (homogeneous) null, that is, the distribution of the  test statistic does not depend on the null distribution of the data. Moreover, the WEC statistic estimates the well-known Henze-Penrose divergence (see Equation \eqref{eq:deltaxy} for the definition) and, consequently, can detect arbitrary differences between two distributions under the homogeneous null. We leverage these properties to obtain a cut-off for the WEC statistic that can detect deviations from the heterogeneous (composite) null, beyond location problems. 
\item For non-asymptotic usage, we develop a novel bootstrap-based calibration procedure for the weighted edge count test statistic (Algorithm \ref{algo:max} in Section \ref{sec:bootstrap}) and use it to test the composite null hypothesis that the two samples arise from the same mixture distribution but with possibly different mixing weights (Equations \eqref{eq:setup-1}--\eqref{eq:setup-3}). Under sample size imbalance, our calibration procedure first explores the larger sample for heterogeneous subgroups. Then it generates an ensemble of mixing weights for the different subgroups and subsequently constructs a collection of surrogate two-samples under the composite null hypothesis. The WEC test statistic is computed for each such surrogate two-samples and %Then, on each of the identified subgroups, it generates a bootstrapped null distribution for the WEC statistic for testing the simple null hypothesis of equality of distributions. 
the ensemble of these WEC test statistics %subgroup-level bootstrapped null distributions
is then used to determine the level $\alpha$ cutoff for testing the composite null hypothesis (Equation \eqref{eq:setup-3}).
\item Our numerical experiments (Section \ref{sec:sims}) reveal that across a wide range of simulation settings, the proposed bootstrap calibration procedure results in a conservative level $\alpha$ cutoff for the WEC test statistic for consistent nonparametric two sample testing involving the composite null hypothesis of Equation \eqref{eq:setup-3}. This is in contrast to existing nonparametric two-sample tests that may lead to incorrect inference in this regime. Furthermore, our empirical evidence suggests that the bootstrap calibration procedure renders the WEC test more powerful than the recently introduced TRUH test \citep{banerjee2020nearest} for the heterogeneous two sample problem. %\red{Should we elaborate a bit more on the empirical findings? Discuss better performance in scale problems?}
\item We apply our proposed hypothesis testing procedure to detect addictive behaviors in online gaming (Section \ref{sec:realdata}). On an anonymized data available from a large video game company in Asia, we test whether players who login after midnight and players who login early in the morning exhibit deviant playing behavior when compared to players with normal gaming behavior. Our analysis reveals that the playing behavior of players who login after midnight is statistically different from the normal gaming behavior, while those who login early, crucially enough, do not exhibit such differences in their playing behavior. These results confirm the findings in extant research that indicate higher tendency towards game addiction for those players who login late \citep{lee2017predictors}. Existing nonparametric two-sample tests, on the other hand, erroneously conclude aberrant gaming behavior for both sets of players, those who login late and those who login early. 
\end{itemize}
%\subsection{Organization}
%The rest of the article is organized as follows. In Section \ref{sec:2} we study the asymptotic properties of the WEC test statistic under our proposed two-sample hypothesis testing framework which is based on the composite null hypothesis that the two samples arise from the same mixture distribution but with possibly different mixing weights (Equations \eqref{eq:setup-1}--\eqref{eq:setup-3}). Section \ref{sec:bootstrap} presents the bootstrap-calibrated WEC test statistic for practical implementation of the WEC test under heterogeneity and the empirical performance of this test is investigated in Section \ref{sec:sims}. In Section \ref{sec:realdata}, we apply our proposed test for detecting addictive player behavior on online video games. The article concludes with a discussion in Section \ref{sec:discuss}. Proofs and other technical details are relegated to the supplement.
%
\section{Nonparametric two-sample testing under heterogeneity}
\label{sec:2}
We begin with a review of the edge count tests in Section \ref{sec:review}. Thereafter, Section \ref{sec:hetero_problem} introduces the heterogeneous two-sample hypothesis testing problem. In Section \ref{sec:theory} we study the asymptotic properties of the WEC statistic for the heterogeneous two-sample problem. %present the proposed bootstrap based procedure for calibrating the WEC test for the heterogeneous two-sample problem. We conclude with a discussion on the implementation of this calibration procedure in Section \ref{sec:implement}.

We first collect some notations that will be used throughout this article. Denote the two independent samples by $\mathcal X_n=\{\bm X_1,\ldots,\bm X_n\}$ and $\mathcal Y_m=\{\bm Y_1,\ldots,\bm Y_m\}$. Suppose each {$\bm X_i\in\mathbb{R}^d$}, $i=1,\ldots,n$, is distributed independently and identically according to a distribution that has cumulative distribution function (cdf) $F_{\bm X}$. Similarly, denote $F_{\bm Y}$ to be the cdf of the distribution of {$\bm Y_j\in\mathbb{R}^d$} for $j=1,\ldots,m$. Denote $N=n+m$ and let %$\mathcal{G}$ be an undirected similarity graph constructed on the pooled observations 
$\mathcal Z_N = \{\mathcal X_n,\mathcal Y_m\}$ be the pooled sample.%, such as a $k-$ Minimum Spanning Tree (MST)\footnote{The $k-$MST is the union of the $1^{st},\ldots,k^{th}$ MSTs, where the $k^{th}$ MST is the spanning tree that minimizes the sum of distances across edges subject to the constraint that this spanning tree does not contain any edge in the $1^{st},\ldots,(k-1)^{th}$ MSTs.} using $L_2$ or $L_1$ distance. We will use $\mathcal{E}_{\mathcal G}$ to denote the edge set of $\mathcal G$. For an edge $e=(i,j)\in\mathcal{E}_{\mathcal G}$, let
%\begin{eqnarray*}
%	J_e = \begin{cases}
%		2,~\text{if observations $i$ and $j$ are from sample $\mathcal Y_m$}\\
%		1,~\text{if observations $i$ and $j$ are from sample $\mathcal X_n$}\\
%		0,~\text{if observations $i$ and $j$ are from different samples}
%	\end{cases}.
%\end{eqnarray*}
%Define
%\begin{equation}
%	\label{eq:r0r1r2}
%	R_k = \sum_{e\in \mathcal E_{\mathcal G}}\mathbb{I}\{J_e=k\},~k=0,1,2.
%\end{equation}
%In Equation \eqref{eq:r0r1r2}, $R_0$ represents the number of between-sample edges, $R_1$ is the number of edges that connect both observations from $\mathcal X_n$ and $R_2$ is the number of edges that connect both observations from $\mathcal Y_m$. 
\subsection{Edge count tests -- a review}
\label{sec:review} 
In their seminal paper \cite{friedman1979multivariate} introduced the edge-count (EC) test for the classical two-sample hypothesis testing problem: 
\begin{align}\label{eq:fxy} 
H_0:F_{\bm X}=F_{\bm Y} \text{ versus } H_1: F_{\bm X}\ne F_{\bm Y}. 
\end{align}
The EC test can be described as follows: 
\begin{itemize} 
\item Construct a similarity graph $\mathcal G$ (based on the pairwise distances between the observations) of the pooled sample $\mathcal Z_N$.
\item Count the number of edges $R_0$ in the graph $\mathcal G$ with one end-point in sample 1 and other in sample 2, and reject $H_0$ in \eqref{eq:fxy} if $R_0$ is `small'. 
\end{itemize}
\cite{friedman1979multivariate} chose $\mathcal G$ to be the $\ell$-minimum spanning tree (MST) of the pooled sample $\mathcal Z_N$ using the $L_2$ distance\footnote{A {\it spanning tree} of a finite set $S \subset \mathbb R^d$ is a connected graph  with vertex-set $S$ and no cycles. A 1-{\it minimum spanning tree} (MST), or simply a MST, of $S$ is a spanning tree which minimizes the sum of distances across the edges of the tree. A  $\ell$-MST of $S$, for $\ell \geq 2$, is the union of the edges in the $\mathrm{(\ell-1)}$-MST together with the edges of the spanning tree that minimizes the sum of distances across edges subject to the constraint that this spanning tree does not contain any edge of the $(\ell-1)$-MST.}. %\footnote{\textcolor{red}{A {\it spanning tree} of a finite set $S \subset \mathbb R^d$ is a connected graph  with vertex-set $S$ and no cycles. A 1-{\it minimum spanning tree} (MST), or simply a MST, of $S$ is a spanning tree which minimizes the sum of distances across the edges of the tree. A  $\ell$-MST of $S$, for $\ell \geq 2$, is the union of the edges in the $\mathrm{(\ell-1)}$-MST together with the edges of the spanning tree that minimizes the sum of distances across edges subject to the constraint that this spanning tree does not contain any edge of the $(\ell-1)$-MST. (BBB: Check definition.)}} 
Thereafter, tests based on other similarity graphs have been proposed. In particular, 
\cite{schilling1986multivariate} and \cite{henze1988multivariate} considered tests where $\mathcal G$ is the nearest-neighbor graph and \cite{rosenbaum} proposed a test where $\mathcal G$ is the minimum non-bipartite matching. The aforementioned tests are all asymptotically distribution-free (the asymptotic distribution of $R_0$ under $H_0$ in \eqref{eq:fxy} does not depend on the distribution of the data), universally consistent (the test has asymptotic power 1 for all alternatives in \eqref{eq:fxy}), and computationally efficient (running time is polynomial in both the number of data points and dimension), making them readily usable in applications. 

Although the EC test is universally consistent (see \citep[Theorem 2]{henze1999multivariate}), \cite{chen2017new} and \cite{chen2018weighted} observed that it has has two major limitations: First, empirical evidence suggest that the EC test has low or no power for scale alternatives even though asymptotically the test is consistent for both location and scale alternatives \citep{henze1999multivariate}. Second, under sample size imbalance the EC test statistic has a relatively large variance and exhibits low power for detecting departures from the null hypothesis in Equation \eqref{eq:fxy}. To mitigate these issues, \cite{chen2017new} and  \cite{chen2018weighted} suggested new tests based on the within sample edges in the similarly graph $\mathcal G$. Towards this suppose $\mathcal{E}_{\mathcal G}$ denotes the edge set of $\mathcal G$. For an edge $e=(i,j)\in \mathcal{E}_{\mathcal G}$, let
\begin{eqnarray*}
J_e = \begin{cases}
	2~\text{if observations $i$ and $j$ are from sample $\mathcal Y_m$,}\\
	1~\text{if observations $i$ and $j$ are from sample $\mathcal X_n$,}\\
	0~\text{if observations $i$ and $j$ are from different samples. }
\end{cases} 
\end{eqnarray*}
For $k \in \{0,1,2\}$, define
\begin{equation}
\label{eq:r0r1r2}
R_k = \sum_{e\in \mathcal E_{\mathcal G}}\mathbb{I}\{J_e=k\} . 
\end{equation}
Note that $R_0$, which counts the number of between-sample edges, is the EC  statistic introduced before. Similarly, $R_1$ is the number of edges with both endpoints in $\mathcal X_n$ and $R_2$ is the number of edges with both endpoints in $\mathcal Y_m$. 

To addresses the issue of low power for scale alternatives,  \cite{chen2017new} 
proposed the Generalized Edge Count (GEC) test, which rejects the null hypothesis $h_0$ in Equation \eqref{eq:fxy} for large values of 
\begin{align}
\label{eq:gec}
\mathcal{R}_g(\mathcal X_n, \mathcal Y_m)=(R_1-\mu_1,R_2-\mu_2)\bm \Sigma^{-1}(R_1,R_2)^T,
\end{align}
where $\mu_k=\mathbb{E}(R_k)$, for $k=1, 2$, and $\bm \Sigma$ is the covariance matrix of $(R_1,R_2)^T$ under the permutation null distribution (see Lemma 3.1 of \cite{chen2017new} for the analytical expression of these quantities). The Weighted Edge Count (WEC) test of \cite{chen2018weighted} addresses the issue of sample-size imbalance by proposing a new test statistic $\mathcal{R}_w(\mathcal X_n,\mathcal Y_m)$ based on the weighted sum of the within sample edges, $R_1$ and $R_2$, as follows:
\begin{align}
\label{eq:wec}
\mathcal{R}_w(\mathcal X_n,\mathcal Y_m)= \frac{1}{N} \left( \dfrac{m}{N}R_1+\dfrac{n}{N}R_2 \right). 
\end{align} 
The weighting scheme controls the variance of $\mathcal{R}_w$ as opposed to the EC test statistic based on $R_0$. The test rejects $H_0$ in \eqref{eq:fxy} for large values of $\mathcal{R}_w(\mathcal X_n,\mathcal Y_m)$, where the cut-off is computed from the permutation or the asymptotic null distribution under $H_0: F_{\bm X}=F_{\bm Y}$. 
Evidence from authors' empirical studies reveal that under sample size imbalance and for location alternatives, the WEC test is more powerful than both the EC test and the GEC test.

\subsection{The heterogeneous two-sample problem}
\label{sec:hetero_problem}
In this section we introduce the heterogeneous two-sample hypothesis testing problem using motivating Example 2 from Section \ref{sec:motivation}. For ease of exposition, we will refer the population of players who exhibit normal playing behavior as the baseline population. So in our example, $\mathcal X_n$ is an i.i.d random sample of size $n$ from the baseline population that has cdf $F_{\bm X}$ and $\mathcal Y_m$ is an i.i.d random sample of size $m$ from the population of current players that has cdf $F_{\bm Y}$. We consider a setting where the heterogeneity in the baseline population is represented by $K$ different subgroups, each having unimodal distributions with distinct modes and cdfs $F_{1}, \ldots, F_K$, such that 
\begin{align}
\label{eq:setup-1}
{F}_{\bm X} = \sum_{a=1}^K w_{a} \, F_{a}, \quad \text{ where } \quad w_{a} \in (0,1) \text{ and } \sum_{a=1}^K w_{a}=1.
\end{align}
Here the number of components $K$, the mixing distributions $F_1,\ldots, F_K$, and the mixing weights $w_1,\ldots, w_K$ are fixed (non-random) but unknown. Furthermore, since $F_{1}, \ldots, F_K$ are cdfs from unimodal distributions with distinct modes, ${F}_{\bm X}$ is well-defined with a unique specification 
i.e,  ${F}_{\bm X} \neq \sum_{a=1}^K \tilde w_{a} \, F_{a}$  if $\tilde w_{a}\neq w_a$ for at least one $a \in \{1,\ldots,K\}$. The population of normal players may exhibit two distinct phenomenon. First, the normal players can have similar playing behavior as the baseline population but a different representation of the $K$ subpopulations than those reflected by the mixing proportions $\{w_1, \ldots, w_K\}$. This may imply that a few baseline subpopulations are completely absent in the population of normal players. Thus, if the normal players do not exhibit a different playing behavior then their cdf $F_{\bm Y}$ lies in a class of distributions $\mathcal{F}(F_{\bm X})$ that contains any convex combination of $\{F_1, \ldots, F_K\}$ including the boundaries, that is, 
\begin{align}
\label{eq:setup-2}
\mathcal{F}(F_{\bm X}) = \left\{Q = \sum_{a=1}^K \lambda_a \, F_a: \lambda_1, \lambda_2, \ldots, \lambda_K \in [0,1] \text{ and } \sum_{a=1}^K \lambda_a=1 \right\}.
\end{align} 
Note that the left and center panels of Figure \ref{fig:motfig_2} are examples of the setting where $F_{\bm Y}\in\mathcal{F}(F_{\bm X})$. Furthermore, $F_{\bm X}\in\mathcal{F}(F_{\bm X})$ in Equation \eqref{eq:setup-2}. Second, if the normal players exhibit a different playing behavior than the baseline population, then $F_{\bm Y}$ would contain at least one non-trivial subpopulation with distribution substantially different from $\{F_1, F_2, \ldots, F_K\}$ or their linear combinations. Then, $F_{\bm Y}\notin \mathcal{F}(F_{\bm X})$ and the right panel of Figure \ref{fig:motfig_2} represents this phenomenon. The heterogeneous two-sample problem that we consider in this article involves testing the following composite null hypothesis: 
\begin{equation}
\label{eq:setup-3}
H_0: F_{\bm Y} \in \mathcal{F}(F_{\bm X}) \quad \text{versus} \quad H_1: F_{\bm Y} \notin \mathcal{F}(F_{\bm X}).
\end{equation}
Existing nonparametric graph-based two-sample tests, such as the EC test, are designed to test the simple null hypothesis $H_0: F_{\bm X}=F_{\bm Y}$, and are not conservative for testing the composite null hypothesis $H_0: F_{\bm Y} \in \mathcal{F}(F_{\bm X})$ of Equation \eqref{eq:setup-3} (see, for example, Proposition 1 in \citep{banerjee2020nearest}). %Next, we discuss a bootstrap based calibration technique for the WEC  test statistic that delivers a conservative procedure for the hypothesis testing problem presented in Equation \eqref{eq:setup-3}.
\subsection{Asymptotic properties of WEC statistic under heterogeneity}
\label{sec:theory}
In this section we will discuss how one can calibrate the WEC statistic under heterogeneity to obtain a test that is asymptotically powerful for general practical alternatives. For this we assume that the baseline cdfs $F_1, F_2, \ldots, F_K$ have unimodal densities $f_1, f_2, \ldots, f_K$ (with respect to Lebesgue measure). Therefore, the baseline population will have density $f_{\bm X} = \sum_{a=1}^K w_af_a$ (recall Equation \eqref{eq:setup-1}), and the set of distributions in Equation \eqref{eq:setup-2}
can be represented in terms of the densities $f_1, f_2, \ldots, f_K$ as: 
\begin{align*}
	%\label{eq:setup-21}
	\left\{ g = \sum_{a=1}^K \lambda_a \, f_a: \lambda_1, \lambda_2, \ldots, \lambda_K \in [0,1] \text{ and } \sum_{a=1}^K \lambda_a=1 \right\} , 
\end{align*} 
and will be denoted by $\mathcal F(f_{\bm X})$. Then, assuming that the population of current players has density $f_{\bm Y}$, the hypothesis testing problem in Equation \eqref{eq:setup-3} can be restated as:  
\begin{equation*}
	%\label{eq:setup-31}
	H_0: f_{\bm Y} \in \mathcal{F}(f_{\bm X}) \quad \text{versus} \quad H_1: f_{\bm Y} \notin \mathcal{F}(f_{\bm X}).
\end{equation*}  
To prove our asymptotic results we assume the following: 
\begin{assumption}\label{assumption:fxy} 
	The densities $f_{\bm X}$ and $f_{\bm Y}$ have a common support $S \subseteq \mathbb{R}^d$ and there exist constants $M_1, M_2 > 0$ such that $M_1 \leq f_{\bm X}(x) \leq M_2$ and $f_{\bm Y}(x) \leq M_2$, for all $x \in S$. 
\end{assumption}
\begin{assumption}\label{assumption:w} 
	The weights of the baseline population are bounded below, that is, there  exists a known constant $L > 0$ such that $w_a > L$, for $1 \leq a \leq K$. 
\end{assumption}
%\textcolor{red}{We also assume that $\hat \gamma_{m, n}$ is a consistent estimate of $\gamma:=\max_{1 \leq a, b \leq K} \delta(f_b, f_a, \rho)$. }

Now, recall the definition of the WEC statistic $\mathcal{R}_w(\mathcal X_n,\mathcal Y_m)$ from Equation \eqref{eq:wec}. Then, we have the following result in the usual asymptotic regime where $m, n \rightarrow \infty$ such that $n/m \rightarrow \rho \in (0, \infty)$. 

\begin{proposition}\label{consistency} Suppose the similarity graph $\mathcal G$ is the $\ell$-MST, for some finite $\ell \geq 1$. Then under Assumptions  \ref{assumption:fxy} and \ref{assumption:w} the following hold:  
	\begin{itemize}
		
		\item[$(1)$]  For any $f_{\bm Y} \in \mathcal{F}(f_{\bm X})$, 
		\begin{equation*}%\label{eq:H0}
			\lim_{m, n \rightarrow \infty}\P_{f_{\bm X}, f_{\bm Y}}\Bigl\{\mathcal{R}_w(\mathcal X_n,\mathcal Y_m) > \gamma\Bigr\} = 0~,
		\end{equation*}
		where  $\gamma:= \frac{\ell  \rho}{ (1+\rho)^2} \left\{ 1 + \frac{(1+\rho)K^2}{L \rho^2} \right\}
		$. 
		
		\item[$(2)$] Suppose $\varepsilon := \left\{ \frac{2 M_2^3 (1+\rho)^3 K^2}{ M_1^2 \rho^3 L} \right\}^{\frac{1}{2}}$. Then 
		\begin{equation*}%\label{eq:H1}
			\lim_{m, n \rightarrow \infty}\P_{f_{\bm X}, f_{\bm Y}}\Bigl\{\mathcal{R}_w(\mathcal X_n,\mathcal Y_m) >  \gamma\Bigr\} =1,
		\end{equation*}
		whenever $f_{\bm Y} \notin \mathcal{F}(f_{\bm X})$ is such that $\int_S \{f_{\bm Y}(x)-f_{\bm X}(x)\}^2  \mathrm dx > \varepsilon^2$. 
	\end{itemize}
\end{proposition}

This result shows that if one chooses the cut-off of the WEC statistic based on $\gamma$, then the probability of Type I error is asymptotic zero for all  $f_{\bm Y} \in \mathcal{F}(f_{\bm X})$. Moreover, the power of the test is asymptotically 1, whenever $f_{\bm Y}$ is $\varepsilon$-far (in the $L_2$-distance) from the baseline density $f_{\bm X}$. In other words, the WEC statistic calibrated as above is consistent whenever the signal strength (measured in terms of the $L_2$-distance) is not too small.  The proof of Proposition \ref{consistency} relies on the following asymptotic result (see \cite[Theorem 4]{chen2018weighted} and \cite[Theorem 2]{henze1999multivariate}): Suppose $\mathcal X_n=\{\bm X_1,\ldots,\bm X_n\}$ and $\mathcal Y_m=\{\bm Y_1,\ldots,\bm Y_m\}$ are i.i.d. samples from from $d$-dimensional distributions with absolutely continuous densities $f_{\bm X}$ and $f_{\bm Y}$, respectively. Then as $m, n \rightarrow \infty$ such that $n/m \rightarrow \rho \in (0, \infty)$, 
\begin{align}\label{eq:Rxynm} 
	\mathcal{R}_w(\mathcal X_n,\mathcal Y_m) \pto \frac{\ell  \rho}{ (1+\rho)^2} \cdot \delta_\rho(f_{\bm X}, f_{\bm Y}), 
\end{align}
where 
\begin{align}\label{eq:deltaxy}
	\delta_\rho(f_{\bm X}, f_{\bm Y}) = \int_S \frac{ \rho f_{\bm X}^2(x) + f_{\bm Y}^2(x)}{( \rho f_{\bm X}(x) + f_{\bm Y}(x))}\mathrm dx .
\end{align}
The quantity $\delta_\rho(f_{\bm X}, f_{\bm Y})$ is known as the {\it Henze-Penrose divergence} between the densities $f_{\bm X}$ and $f_{\bm Y}$ and plays a central role in the consistency of edge-count type tests. The proof of Proposition \ref{consistency} entails showing that $\delta_\rho(f_{\bm X}, f_{\bm Y}) $ is uniformly bounded above under the composite null $H_0: f_{\bm Y} \in \mathcal F({f_{\bm X}})$ and is `large' under the alternative (details are given in Section \ref{sec:pfconsistency} of the Supplement). 
\begin{remark} Note that Assumption \ref{assumption:fxy} requires that the baseline density $f_{\bm X}$ is bounded below in its support. This is a well-known assumption that is often required in deriving asymptotic properties of methods based on geometric graphs and nearest-neighbors (see, for example, \cite{biau2015lectures,py,zhao2022analysis} and the references therein). Although this assumption technically rules out some natural distributions, from a practical standpoint, there is no real concern because one can approximate the univariate density by truncating it to a large interval, on which the above result applies. Moreover, it is possible to remove this assumption by replacing the $L_2$ separation condition in Proposition \ref{consistency} with a truncated $L_2$ separation condition (see Remark \ref{remark:assumption} in the supplement for details). 
	%However, this still does not apply for $d=1$, where it is necessary to assume the compactness of the support, in order to ensure that the limit in \eqref{eq:TNM_limit} is finite.  This is a well-known constraint which arises in a large family of random geometric graphs, while dealing with the asymptotics of edge-lengths (see, for example, \cite[Theorem 1.1]{py} and the references therein). Even though the compactness assumption technically rules out some natural distributions, from a practical standpoint, there is no real concern because one can approximate the univariate density by truncating it to a large interval, on which the above result applies. Incidentally, there has been recent work on relaxing the compactness and density bounded below  assumptions in the related problems of nearest-neighbor classification \cite{cannings2017local,gadat2016classification} and entropy estimation \cite{berrett2019efficient}, which could provide useful insights on how to relax these assumptions from Theorem \ref{THM:EXPTNM}, and what are the effects of tail behavior on the heterogeneity testing problem. 
\end{remark}
\section{Bootstrap based calibration of the {WEC} test statistic}
\label{sec:bootstrap}
Here we develop a bootstrap based re-calibration procedure for practical implementation of the WEC test under heterogeneity. Under sample size imbalance, our proposed calibration procedure for the WEC test statistic first explores the larger sample $\mathcal X_n$ for heterogeneous subgroups. %Then, on each of the identified subgroups it generates a bootstrapped null distribution for the WEC statistic for testing the simple null hypothesis $H_0:F_{\bm X}=F_{\bm Y}$ against the alternative $H_1:F_{\bm X}\ne F_{\bm Y}$. The ensemble of the subgroup-level bootstrapped null distributions is then used to determine the level $\alpha$ cutoff for testing the composite null hypothesis of Equation \eqref{eq:setup-3}.
To determine the number of such subgroups $K$ in $\mathcal X_{n}$, we use the prediction strength approach of \citet{tibshirani2005cluster}, which gives an estimate $\hat K$ of $K$. The class membership of the baseline samples $\mathcal X_n$ is then determined using a $\hat K$-means algorithm. Denote by $\hat J_a \subseteq \{1, 2, \ldots, n\}$ to be the subset of indices estimated to be in class $a$ by the $\hat K$-means algorithm where $1 \leq a \leq \hat K$. Let $n_a=|\hat{J}_a|$ to be the cardinality of class  $a$ and denote $\mathcal X_{\hat J_a}=\{\bm X_i: i \in \hat J_a \}$ to be the corresponding  subset of the baseline samples estimated to be in class $a$. Note that $\mathcal{X}_n=\{\mathcal{X}_{\hat J_a}: a=1, 2, \ldots,\hat{K}\}$ and $\sum_{a=1}^{\hat{K}} n_a=n$.  
Our calibration procedure then repeats the following three steps a large number of times: 
\begin{enumerate}
\item Mixing proportions are randomly sampled from the $\hat K$-dimensional simplex $\mathcal{S}_{\hat K}$ where
$\mathcal{S}_{\hat K}=\{(\lambda_1, \ldots, \lambda_{\hat K}) \in \mathbb{R}^{\hat K}: \lambda_a \in [0, 1], \text{ for } 1 \leq a \leq \hat{K}, \text{ and } \sum_{a=1}^{\hat K} \lambda_a=1\}.$
\item For a given realization of the mixing proportions from $\mathcal{S}_{\hat K}$, surrogate baseline and normal players samples are generated from $\mathcal{F}(F_{\bm X})$.
\item The WEC test statistic is computed using the two samples obtained from step (2).
\end{enumerate}
We now discuss these three steps below.

For step (1), and for each $b=1,\ldots,B$, denote $(\lambda_1^{(b)}, \ldots, \lambda_{\hat{K}}^{(b)})$ to be a random sample from $\mathcal{S}_{\hat K}$. Given the mixing weights $\{\lambda_1^{(b)},$ $\ldots, \lambda_{\hat K}^{(b)}\}$, step (2) involves constructing surrogate baseline and normal players samples from $\mathcal{F}(F_{\bm X})$ as follows: for each $a\in\{1,\ldots,\hat{K}\}$,  randomly sample $\lceil m\lambda_a^{(b)} \rceil$ elements without replacement from $\mathcal{X}_{\hat J_a}$. Then the chosen elements constitute the surrogate normal players sample in class $a$ and are denoted by $\mathbb{Y}^{(b)}_a=\{\bm X^{(b)}_1,\ldots, \bm X^{(b)}_{\lceil n\lambda_a^{(b)} \rceil}\}$. The remaining $n_a-\lceil m\lambda_a^{(b)} \rceil$ elements in $\mathcal{X}_{\hat J_a}$ form the residual baseline sample $\mathbb{X}^{(b)}_{a}$ in class $a$. We combine these samples over the $\hat K$ classes to get the surrogate normal players sample as $\mathcal{Y}^{(b)}_{m}=\{\mathbb{Y}^{(b)}_{a}: a=1,\ldots,\hat{K}\}$ and the corresponding baseline sample as $\mathcal{X}^{(b)}_{\tilde{n}}=\{\mathbb{X}^{(b)}_{a}: a=1,\ldots,\hat{K}\}$, where $\tilde{n}=\sum_{a=1}^{\hat{K}}(n_a-\lceil m\lambda^{(b)}_a \rceil)$. Finally, for step (3) the bootstrapped samples in the {$b^{th}$} round, $\mathcal{X}^{(b)}_{\tilde{n}}$ and $\mathcal{Y}^{(b)}_{m}$ (which are surrogates for $\mathcal X_n$ and $\mathcal Y_m$, respectively), can be used to compute the WEC test statistic $\mathcal{R}_w^{(b)}\coloneqq\mathcal{R}_w(\mathcal{X}^{(b)}_{\tilde{n}},\mathcal{Y}^{(b)}_{m})$. 

The bootstrap calibration procedure described above is summarized in Algorithm \ref{algo:max}. % and the same algorithm provides the level $\alpha$ cutoff using the GEC test statistic when $\mathcal{R}_w(\bm{X}^{(b)}_{\tilde{n}},\bm{Y}^{(b)}_{m})$ is replaced by $\mathcal{R}_g(\bm{X}^{(b)}_{\tilde{n}},\bm{Y}^{(b)}_{m})$ in Algorithm \ref{algo:max}.
\begin{algorithm}[!h]
\KwIn{\small The parameters $n$ and $\alpha$. The baseline samples $\mathcal{X}_n$, and the estimates $\hat{K}$ and $\{\hat J_a : a=1,\ldots,\hat{K}\}$ from the $K$-means algorithm with $n_a=|\hat{J}_a|$.}
%\Parameter{$n$, $\tau_{fc}$, $\alpha$.}
\KwOut{\small The bootstrapped level $\alpha$ cutoff $r_{n,m,\alpha}$.}
%initialization\;
\For{$b=1,\ldots,B$}{\small
	STEP 1: Random sample $\{\lambda_1^{(b)}, \ldots, \lambda_{\hat K}^{(b)}\}$ from the $\hat K$-dimensional simplex\;
	\For{$a=1,\ldots,\hat{K}$}{
		
		\eIf{$\lceil m\lambda_a^{(b)} \rceil \le{n}_a$}{
			STEP 2: Draw a simple random sample 
			$\mathbb{Y}_a^{(b)}=\{\bm X^{(b)}_1, \ldots, \bm X^{(b)}_{\lceil m\lambda_a^{(b)} \rceil}\}$ without replacement from $\mathcal{X}_{\hat J_a}$\;
			STEP 3: $\mathbb{X}_a^{(b)}=\mathcal{X}_{\hat J_a} \backslash \mathbb{Y}_a^{(b)}$ the baseline residual sample in class $a$\;				
		}{
			Stop: Go to STEP 1\;				
		}
	}
	Surrogate normal players sample: $\mathcal{Y}^{(b)}_{m}=\{\mathbb{Y}^{(b)}_{a}: a=1,\ldots,\hat{K}\}$\;
	Baseline sample: $\mathcal{X}^{(b)}_{\tilde{n}}=\{\mathbb{X}^{(b)}_{a}: a=1,\ldots,\hat{K}\}$\;
	STEP 4: Calculate $\mathcal{R}_w^{(b)}\coloneqq\mathcal{R}_w(\mathcal{X}^{(b)}_{\tilde{n}},\mathcal{Y}^{(b)}_{m})$\;
}
\small STEP 5: Return $r_{n,m,\alpha}=\min\{\mathcal{R}_w^{(b)}: \frac{1}{B} \sum_{r=1}^{B}\bm 1 \{\mathcal{R}_w^{(r)}\ge \mathcal{R}_w^{(b)}\} \le \alpha\}$.
\caption{\small Bootstrap cut-off for a level $\alpha$ test using $\mathcal{R}_w(\mathcal X_n,\mathcal Y_m)$}
\label{algo:max}
\end{algorithm}
The computational complexity of our calibration procedure depends on two key steps: (i) computation of the estimated number of clusters $\hat{K}$, and (ii) computation of the WEC test statistic over $B$ bootstrap samples. To estimate $K$, we use prediction strength along with a $K-$means algorithm where the target number of clusters and the maximum
number of iterations over which the $K-$means algorithm runs before stopping are both fixed. Thus step (i) has $O(nd)$ complexity. The calculations in step (ii) can be distributed across the $B$ bootstrap samples but for each $b\in\{1,\ldots,B\}$ computation of the WEC test statistic requires calculating the $N\times N$ distance matrix and constructing the MST on the pooled sample $\mathcal Z_N$. The former has $O(n^2d)$ complexity while the latter has $O((n-1)\log n)$ complexity. Therefore, the overall computational complexity of Algorithm \ref{algo:max} is $O(n^2d)$.
% is  , the computational cost of estimating $\mathcal{R}_w(\bm{X}^{(b)}_{\tilde{n}},\bm{Y}^{(b)}_{m})$
%for a fixed $b$ is \trf{$O(?)$}. Therefore the overall computational complexity of Algorithm \ref{algo:max} is $O(?)$. \trf{Distance matrix is $O(n^2d)$ and the MST is $O(|\mathcal{G}|\log n)$ where $|\mathcal{G}|=n-1$ is the number of edges in the MST $\mathcal{G}$.}
\subsection{Implementation}
\label{sec:implement}
Here we discuss several aspects related to the implementation of the proposed calibration procedure presented in Algorithm \ref{algo:max}. First, for the numerical experiments and real data analysis (Sections \ref{sec:sims} and \ref{sec:realdata}, respectively), we set $B=200$ and implement a version of Algorithm \ref{algo:max} where we sample the mixing proportions $\{\lambda_1, \ldots, \lambda_{\hat K}\}$ only from the corners of $\mathcal{S}_{\hat{K}}$, thus placing most weight on the corners of $\mathcal{S}_{\hat{K}}$. %Tshis sampling scheme ensures that the mechanism for generating the mixing proportions places most weight on the corners of $\mathcal{S}_{\hat{K}}$.
Second, while Algorithm \ref{algo:max} provides a calibration approach for the WEC test statistic, it can also be used to obtain the level $\alpha$ cutoff for the GEC test statistic for testing the composite null hypothesis of Equation \eqref{eq:setup-3}. To do that, we replace $\mathcal{R}_w(\mathcal{X}^{(b)}_{\tilde{n}},\mathcal{Y}^{(b)}_{m})$ in Algorithm \ref{algo:max} with $\mathcal{R}_g(\mathcal{X}^{(b)}_{\tilde{n}},\mathcal{Y}^{(b)}_{m})$ where GEC test statistic is defined as
$\mathcal{R}_g(\mathcal X_n,\mathcal Y_m)=(R_1-\mu_1,R_2-\mu_2)\bm \Sigma^{-1}(R_1,R_2)^T,
$
with $\mu_k=\mathbb{E}(R_k),~k=1,2$ and $\bm \Sigma$ being the covariance matrix of $(R_1,R_2)^T$ under the permutation null distribution (See Lemma 3.1 of \cite{chen2017new} for the analytical expression of these quantities). In Sections \ref{sec:sims} and \ref{sec:realdata} we report the performance of both WEC and GEC test statistics using the bootstrap calibration procedure presented in Algorithm \ref{algo:max}. Third, we rely on the R-package {\sf gTests} to calculate the WEC and GEC statistics using $5-$MST on the pooled sample, which is a recommended practical choice \citep{chen2017new}.

\section{Numerical experiments}
\label{sec:sims}
In this section we assess the numerical performance of the bootstrap calibrated WEC (BWEC) and GEC (BGEC) tests against the following four competing testing procedures across a wide range of simulation settings: (i) Edgecount (EC) test, (ii) Generalized edgecount (GEC) test, (iii) Weighted edgecount (WEC) test, and (iv) TRUH test of \cite{banerjee2020nearest}. For implementing the three edge count tests, we rely on the R package \texttt{gtests} and while for TRUH, we use the code available at \citet{banerjee2020nearest} with the default specification of $\tau_{fc}=1$. Note that amongst the aforementioned four competing tests, only the TRUH statistic is designed to test the composite null hypothesis of Equation \eqref{eq:setup-3}, while the three variants of the edge count test were developed to test the simple null hypothesis $H_0: F_{\bm X}=F_{\bm Y}$ against the alternative $H_1:F_{\bm X}\ne F_{\bm Y}$.

In our numerical experiments, we simulate $\mathcal X_n$ and $\mathcal Y_m$ from $F_{\bm X}$ and $F_{\bm Y}$, respectively, and for each testing procedure, we report the proportion of rejections across $500$ repetitions of the composite null hypothesis test $H_0: F_{\bm Y}\in\mathcal{F}(F_{\bm X})~vs~H_1:F_{\bm Y}\notin \mathcal{F}(F_{\bm X})$ at $5\%$ level of significance. The R code that reproduces our simulation results is available at \url{https://www.dropbox.com/sh/fstb0yzv7yjnfgc/AAA9SbY9trfsTpEmgGEiA6Yva?dl=0}. 
\subsection{Experiment 1}
\label{sec:simexp1}
We consider a setting where $F_{\bm X}$ is the cdf of a $d-$dimensional Gaussian mixture distribution with three components:
$F_{\bm X}=0.3\mathcal{N}_d(\bm\mu_1,\bm \Sigma_1)+0.3\mathcal{N}_d(\bm\mu_2,\bm \Sigma_2)+0.4\mathcal{N}_d(\bm\mu_3,\bm \Sigma_3).
$
Here $\bm \mu_1=\bm 0_{d},~\bm \mu_2=-3\bm 1_{d},~\bm\mu_3=-\bm{\mu}_2$, and $\bm \Sigma_K$, for $K=1,2,3$, are $d\times d$ symmetric positive definite matrices with eigenvalues randomly generated from the interval $[1,~10]$. We consider two scenarios for simulating $\mathcal Y_m$ from $F_{\bm Y}$. In Scenario I we let $F_{\bm Y}=0.1\,\mathcal{N}_d(\bm\mu_1,\bm \Sigma_1)+0.1\,\mathcal{N}_d(\bm\mu_2,\bm \Sigma_2)+0.8\,\mathcal{N}_d(\bm\mu_3,\bm \Sigma_3)$. Thus, $F_{\bm Y}\in\mathcal{F}(F_{\bm X})$ since $F_{\bm Y}$ has all the subpopulations present in $F_{\bm X}$ but at different proportions. For Scenario II we consider $F_{\bm Y}=0.1\,\mathcal{N}_d(\bm\mu_1,\bm \Sigma_1)+0.1\,\mathcal{N}_d(\bm\mu_2,\bm \Sigma_2)+0.8\,\mathcal{N}_d(\bm\mu_3,0.25\bm \Sigma_3)$. The third component in the above mixture differs with respect to its scale when compared to the third component of $F_{\bm X}$. So this setting presents a scenario where $F_{\bm Y}\notin\mathcal{F}(F_{\bm X})$ and the composite null $H_0$ is not true.
\begin{table}[htbp]
\renewcommand{\arraystretch}{0.6}
\centering
\caption{\small Rejection rates at $5\%$ level of significance: Experiment 1 and Scenario 1 wherein $H_0: F_{\bm Y} \in \mathcal{F}(F_{\bm X})$ is true.}
\begin{tabular}{lcccccc}
	\hline
	& \multicolumn{3}{c}{$n = 500,~m = 50$} & \multicolumn{3}{c}{$n = 2000,~m = 200$} \\
	\hline
	\multicolumn{1}{c}{Method} & $d=5$   & $d=15$  & $d=30$  & $d=5$   & $d=15$  & $d=30$ \\
	\hline
	\texttt{EC} test & 0.224 & 0.144 & 0.134 & 0.710 & 0.364 & 0.270 \\
	\texttt{GEC} test & 0.568 & 0.592 & 0.618 & 0.984 & 0.986 & 0.992 \\
	\texttt{WEC} test & 0.724 & 0.738 & 0.750 & 0.998 & 0.994 & 0.998 \\
	\hline
	\texttt{TRUH} test  & 0.028 & 0.030 & 0.020 & 0.020 & 0.018 & 0.008 \\
	\hline
	\texttt{BGEC} test & 0.000 & 0.000 & {0.038} & 0.000 & 0.000 & 0.010 \\
	\texttt{BWEC} test & 0.000 & 0.000 & {0.040} & 0.000 & 0.000 & 0.010 \\
	\hline
\end{tabular}%
\label{tab:exp1set1}%
\end{table}%
\begin{table}[htbp]
\renewcommand{\arraystretch}{0.6}
\centering
\caption{{\small Rejection rates at $5\%$ level of significance: Experiment 1 and Scenario 2 wherein $H_0: F_{\bm Y} \in \mathcal{F}(F_{\bm X})$ is false}.}
\begin{tabular}{lcccccc}
	\hline
	& \multicolumn{3}{c}{$n = 500,~m = 50$} & \multicolumn{3}{c}{$n = 2000,~m = 200$} \\
	\hline
	\multicolumn{1}{c}{Method} & $d=5$   & $d=15$  & $d=30$  & $d=5$   & $d=15$  & $d=30$ \\
	\hline
	\texttt{EC} test & 0.310 & 0.000 & 0.000 & 1.000 & 0.000 & 0.000 \\
	\texttt{GEC} test & 1.000 &  1.000 & 1.000 & 1.000 & 1.000 & 1.000 \\
	\texttt{WEC} test & 1.000 &  1.000 & 1.000 & 1.000 & 1.000 & 1.000 \\
	\hline
	\texttt{TRUH} test  & 0.002 & 0.000 & 0.000 & 0.000 & 0.000 & 0.000 \\
	\hline
	\texttt{BGEC} test & 0.966 & 1.000 & 0.996 & 0.998 & 1.000 & 1.000 \\
	\texttt{BWEC} test & 0.918 & 1.000 & 0.996 & 0.998 & 1.000 & 1.000 \\
	\hline
\end{tabular}%
\label{tab:exp1set2}%
\end{table}%

Table \ref{tab:exp1set1} reports the rejection rates for Scenario I for varying $(n,m,d)$. We note that \texttt{TRUH}, \texttt{BWEC} and \texttt{BGEC} return rejection rates that are below the prespecified $0.05$ level and are conservative across the six testing scenarios considered in Table \ref{tab:exp1set1}. The three edge count tests, on the other hand, have substantially higher rejection rates. This is not surprising as the three edge count statistics are designed to test the simple null hypothesis $F_{\bm X}=F_{\bm Y}$ as opposed to the composite null hypothesis of Equation \eqref{eq:setup-3}. For Scenario II, the rejection rates are reported in Table \ref{tab:exp1set2} and they reveal that with the exception of \texttt{TRUH} and the \texttt{EC} test, all other competing testing procedures are powerful in detecting departures from $H_0: F_{\bm Y} \in \mathcal{F}(F_{\bm X})$. When $d$ is moderately high, the \texttt{EC} test, in particular, is known to have low power under sample size imbalance and the presence of scale alternatives exacerbates this problem. The \texttt{WEC} and \texttt{GEC} tests are designed to address these weaknesses of the \texttt{EC} test, and from Table \ref{tab:exp1set2} we observe substantially higher power for these two tests compared to the original \texttt{EC} test. The \texttt{BWEC} and \texttt{BGEC} tests, while conservative, continue to be powerful in detecting departures from the composite null hypothesis of Equation \eqref{eq:setup-3}. 

The performance of \texttt{TRUH} test under scenarios I and II is worth noting. From tables \ref{tab:exp1set1} and \ref{tab:exp1set2}, we observe that while \texttt{TRUH} is conservative for testing $H_0: F_{\bm Y} \in \mathcal{F}(F_{\bm X})$ versus $H_1: F_{\bm Y} \notin \mathcal{F}(F_{\bm X})$, it is considerably less powerful than \texttt{BGEC} and \texttt{BWEC} tests for detecting departures from $H_0$. In fact, across all our simulation settings \texttt{TRUH}, while conservative, is relatively less powerful than \texttt{BGEC} and \texttt{BWEC} tests. Such a behavior of \texttt{TRUH} is potentially due to its inability to detect departures from $H_0$ when the components of $F_{\bm Y}$ and $F_{\bm X}$ differ only with respect to their scales. %For scenario I, we see that the bootstrapped calibrated Edge Count tests denoted \texttt{B\_GE Count} and \texttt{B\_WE Count} return the smallest rejection rates along with \texttt{TRUH}.  For scenario II, however, \texttt{TRUH} fails to detect $G\notin \mathcal{F}(F_0)$ while \texttt{B\_GE Count} and \texttt{B\_WE Count} correctly detect $G\notin \mathcal{F}(F_0)$.  

\subsection{Experiment 2}
\label{sec:simexp2}
In the setting of Experiment 2, we let $F_{\bm X}$ and $F_{\bm Y}$ to include non-Gaussian components. So we let 
$F_{\bm X}=0.5~{\mathrm{Gam}}_d(\text{shape}=5\bm 1_d,\text{rate}=\bm 1_d,\bm \Sigma_1)+0.5~{\mathrm{Exp}}_d(\text{rate}=\bm 1_d,\bm \Sigma_2),$
where ${\mathrm{Gam}}_d$ and ${\mathrm{Exp}}_d$ are $d-$dimensional Gamma and Exponential distributions. The multivariate Gamma and Exponential distributions are constructed using a Gaussian copula and the R-package \texttt{lcmix} \citep{lcmix,xue2000multivariate} allows sampling from these distributions. 
The correlation matrices $\bm \Sigma_1$ and $\bm \Sigma_2$ are tapering matrices with positive and negative autocorrelations as follows: $(\bm \Sigma_1)_{ij}=0.7^{|i-j|}$ and $(\bm \Sigma_2)_{ij}=-0.9^{|i-j|}$ for $1\le i,j\le d$.
For simulating $\mathcal Y_n$ from $F_{\bm Y}$, we consider two scenarios. In Scenario I, $F_{\bm Y}=0.05~{\mathrm{Gam}}_d(\text{shape}=5\bm 1_d,\text{rate}=\bm 1_d,\bm \Sigma_1)+0.95~{\mathrm{Exp}}_d(\text{rate}=\bm 1_d,\bm \Sigma_2)$. In this setting $F_{\bm Y}$ has both the components of $F_{\bm X}$ but with different mixing proportions. So $H_0: F_{\bm Y}\in\mathcal{F}(F_{\bm X})$ is true, however this is a particularly challenging setting as a majority of the samples from $F_{\bm Y}$ arise from only one of the components of $F_{\bm X}$. In Scenario II, $F_{\bm Y}=0.8~{\mathrm{Gam}}_d(\text{shape}=5\bm 1_d,\text{rate}=\bm 1_d,\bm \Sigma_1)+0.2~{\mathrm{Exp}}_d(\text{rate}=1.5\bm 1_d,0.25\bm \Sigma_3)$ where $(\bm \Sigma_3)_{ij}=0.9^{|i-j|}$. In this setting, the second component of $F_{\bm Y}$ differs from the second component of $F_{\bm X}$ with respect to their correlation matrices, and so $H_0:F_{\bm Y}\in\mathcal{F}(F_{\bm X})$ is not true. 
\begin{table}[!h]
\renewcommand{\arraystretch}{0.6}
\centering
\caption{\small Rejection rates at $5\%$ level of significance: Experiment 2 and Scenario 1 wherein $H_0: F_{\bm Y} \in \mathcal{F}(F_{\bm X})$ is true.}
\begin{tabular}{lcccccc}
	\hline
	& \multicolumn{3}{c}{$n = 500,~m = 50$} & \multicolumn{3}{c}{$n = 2000,~m = 200$} \\
	\hline
	\multicolumn{1}{c}{Method} & $d=5$   & $d=15$  & $d=30$  & $d=5$   & $d=15$  & $d=30$ \\
	\hline
	\texttt{EC} test & 0.418 & 0.312 & 0.258 & 0.944 & 0.798 & 0.684 \\
	\texttt{GEC} test & 0.812 & 0.810 & 0.774 & 1.000 & 1.000 & 1.000 \\
	\texttt{WEC} test & 0.928 & 0.908 & 0.912 & 1.000 & 1.000 & 1.000 \\
	\hline
	\texttt{TRUH} test  & 0.000 & 0.000 & 0.000 & 0.000 & 0.000 & 0.000 \\
	\hline
	\texttt{BGEC} test & 0.002 & 0.004 & 0.000 & 0.000 & 0.000 & 0.000 \\
	\texttt{BWEC} test & 0.000 & 0.004 & 0.000 & 0.000 & 0.000 & 0.000 \\
	\hline
\end{tabular}%
\label{tab:exp2set1}%
\end{table}%

Table \ref{tab:exp2set1} reports the rejection rates for Scenario I. We find that both \texttt{TRUH} and the bootstrapped WEC and GEC tests support $H_0:F_{\bm Y}\in\mathcal{F}(F_{\bm X})$ while the three edge count tests overwhelmingly reject $H_0$, which is an incorrect decision under Scenario I. 
\begin{table}[!h]
\renewcommand{\arraystretch}{0.6}
\centering
\caption{\small Rejection rates at $5\%$ level of significance: Experiment 2 and Scenario 2 wherein $H_0: F_{\bm Y} \in \mathcal{F}(F_{\bm X})$ is false.}
\begin{tabular}{lcccccc}
	\hline
	& \multicolumn{3}{c}{$n = 500,~m = 25$} & \multicolumn{3}{c}{$n = 2000,~m = 100$} \\
	\hline
	\multicolumn{1}{c}{Method} & $d=5$   & $d=15$  & $d=30$  & $d=5$   & $d=15$  & $d=30$ \\
	\hline
	\texttt{EC} test & 0.424 & 0.250 & 0.096 & 0.978 & 0.926 & 0.806 \\
	\texttt{GEC} test & 0.724 &  0.822 & 0.838 & 1.000 & 1.000 & 1.000 \\
	\texttt{WEC} test & 0.824 &  0.886 & 0.904 & 1.000 & 1.000 & 1.000 \\
	\hline
	\texttt{TRUH} test & 0.050 & 0.040 & 0.036 & 0.026 & 0.050 & 0.024 \\
	\hline
	\texttt{BGEC} test & 0.188 & 0.332 & 0.398 & 0.190 & 0.990 & 1.000 \\
	\texttt{BWEC} test & 0.194 & 0.340 & 0.378 & 0.188 & 0.990 & 1.000 \\
	\hline
\end{tabular}%
\label{tab:exp2set2}%
\end{table}%
Table \ref{tab:exp2set2} reports the rejection rates for Scenario II when the sample size imbalance is $0.05$ as opposed to $0.1$ in all of the earlier settings. Detecting departures from $H_0$ under such imbalance is a difficult task as a majority of the samples from $F_{\bm Y}$ arise from the first component of $F_{\bm X}$ and the competing testing procedures must rely on a few observations to reject $H_0$ in Scenario II. Table \ref{tab:exp2set2} reveals that the rejection rates of the three edge count tests and the proposed \texttt{BWEC}, \texttt{BGEC} tests are substantially higher than \texttt{TRUH}. While the edge count tests have the highest rejection rates across both scenarios I and II, the rejection rates of \texttt{BWEC}, \texttt{BGEC} tests improve as the sample size increases in Table \ref{tab:exp2set2}. The two scenarios under Experiment 2 reveal that the bootstrapped calibrated WEC and GEC tests are conservative than edge count tests for testing the composite null hypothesis $H_0: F_{\bm Y} \in \mathcal{F}(F_{\bm X})$ and more powerful than \texttt{TRUH} for detecting departures from $H_0$. 

In Section \ref{app:exp3} of the supplement we report the performance of \texttt{BWEC}, \texttt{BGEC} tests under an additional setting where the samples $\mathcal X_n$ and $\mathcal Y_m$ exhibit zero inflation across the $d$ dimensions.
\section{Detecting player addiction in online video games}
\label{sec:realdata} 
Online video game addiction is a phenomenon wherein a small subgroup of players are involved in excessive and compulsive use of these games that may ultimately result in social and/or emotional problems \citep{lemmens2009development}. Infact, game addiction was included as a disorder in the Diagnostic
and Statistical Manual for Mental Disorders (see DSM-5-TR from the American Psychiatric Association\footnote{https://psychiatry.org/psychiatrists/practice/dsm}) because of an increased risk of clinically significant problems associated with online gaming \citep{petry2014international}. Therefore, for game managers identifying and regulating addicted players is critical because incorrectly rewarding addiction via promotions may lead to high reputation risk for the gaming platform. 

Extant research finds that players who login late at night exhibit a higher tendency towards game addiction \citep{lee2017predictors} and until recently South Korea had prohibited young players from playing online video games between midnight and 6:00 AM. In this application, we rely on an anonymized data available from a large video game company in Asia to test whether players who login after midnight exhibit deviant playing behavior when compared to players with baseline gaming behavior. Our data hold player level information for $d=16$ playing characteristics, such as player's game level, number of friends that they have, number of strategic missions that they completed in the game, etc across $7$ days (see Table \ref{app:tab_datadic} in Supplementary Section \ref{app:data} for a description of these characteristics). For each day, we have access to the following three samples; a sample of players who login post midnight (\texttt{Late}) $\mathcal Y_1=\{\bm Y_{11},\ldots,\bm Y_{1m_1}\}$, an independent sample of players who login between 8 AM - 9 AM local time (\texttt{Early}), $\mathcal Y_2=\{\bm Y_{21},\ldots,\bm Y_{2m_2}\}$ and an independent sample of players who exhibit normal playing behavior (\texttt{Baseline}) $\mathcal X_n=\{\bm X_{1},\ldots,\bm X_{n}\}$. Suppose $\mathcal Y_1$ are a random sample from a distribution with CDF $F^{(1)}_{\bm Y_1}$ and $\mathcal Y_2$ are a random sample from a distribution with CDF $F^{(2)}_{\bm Y_2}$. We consider the following two hypothesis testing problems: (i) whether the playing behavior of \texttt{Late} players is different from \texttt{Baseline} players, $H_{01}: F^{(1)}_{\bm Y_1}\in\mathcal{F}(F_{\bm X})~vs~H_{11}: F^{(1)}_{\bm Y_1}\notin \mathcal{F}(F_{\bm X})$, and (ii) whether the playing behavior of \texttt{Early} players is different from \texttt{Baseline} players, $H_{02}: F^{(2)}_{\bm Y_2}\in\mathcal{F}(F_{\bm X})~vs~H_{12}: F^{(2)}_{\bm Y_2}\notin \mathcal{F}(F_{\bm X})$. %For each day, we have access to the following three samples; a sample of addicted players (\texttt{Addicted}) $\mathcal Y_1=\{\bm Y_{11},\ldots,\bm Y_{1m_1}\}$, an independent sample of engaged players who are \texttt{Early Starters}, i.e they login between 8 AM - 9 AM local time, $\mathcal Y_2=\{\bm Y_{21},\ldots,\bm Y_{2m_2}\}$ and an independent sample of engaged players who login during the rest of the day, i.e, after 9 AM (\texttt{All Others}) $\mathcal X_n=\{\bm X_{1},\ldots,\bm X_{n}\}$. Suppose $\mathcal Y_1$ are a random sample from a distribution with CDF $F^{(1)}_{\bm Y_1}$ and $\mathcal Y_2$ are a random sample from a distribution with CDF $F^{(2)}_{\bm Y_2}$. We consider the following two hypothesis testing problems: (i) whether the playing behavior of \texttt{Addicted} players is different from \texttt{All Others}, $H_{01}: F^{(1)}_{\bm Y_1}\in\mathcal{F}(F_{\bm X})~vs~H_{11}: F^{(1)}_{\bm Y_1}\notin \mathcal{F}(F_{\bm X})$, and (ii) whether the playing behavior of \texttt{Early Starters} is different from \texttt{All Others}, $H_{02}: F^{(2)}_{\bm Y_2}\in\mathcal{F}(F_{\bm X})~vs~H_{12}: F^{(2)}_{\bm Y_2}\notin \mathcal{F}(F_{\bm X})$. 
\begin{figure}[!h]
\centering
\includegraphics[width=1\linewidth]{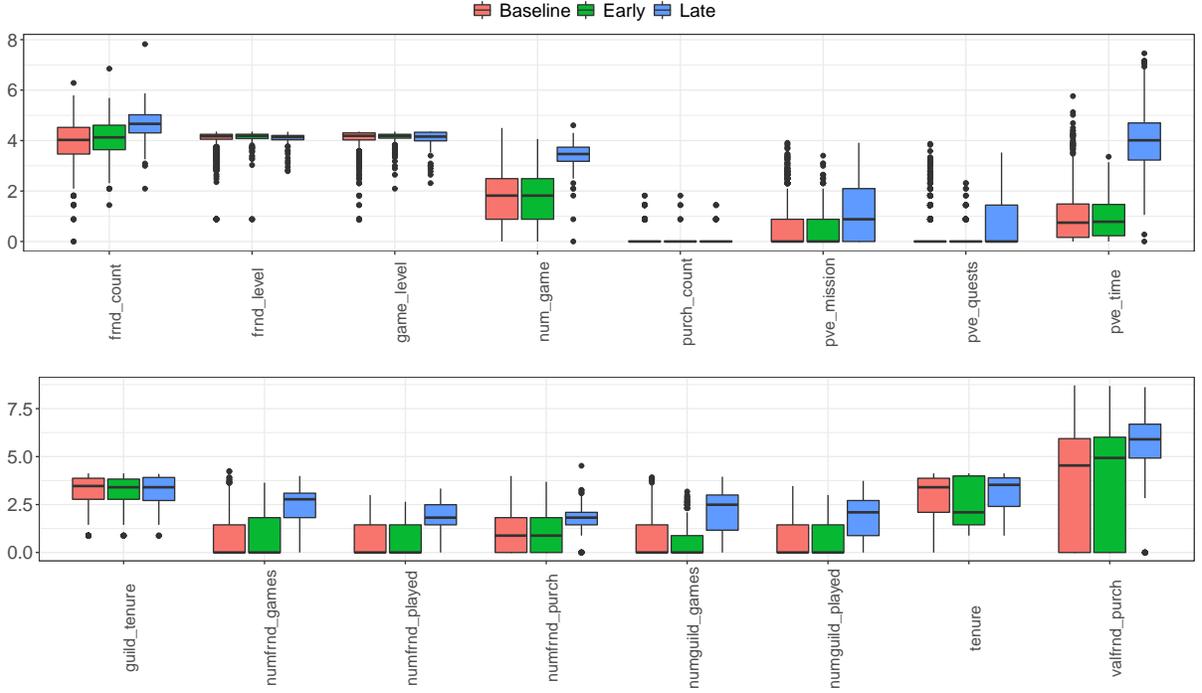}
\caption{\small Box plot of the $16$ playing characteristics on day 5. Data are \texttt{arcsinh} transformed and the variable \texttt{pve\_time} is reported in hours. Here $m_1=143$, $m_2=216$ and $n=2,340$. See Table \ref{app:tab_datadic} in Supplementary Section \ref{app:data} for a description of these characteristics.}
\label{fig:realdata_boxplot}
\end{figure}

\begin{figure}[!h]
\centering
\includegraphics[width=0.8\linewidth]{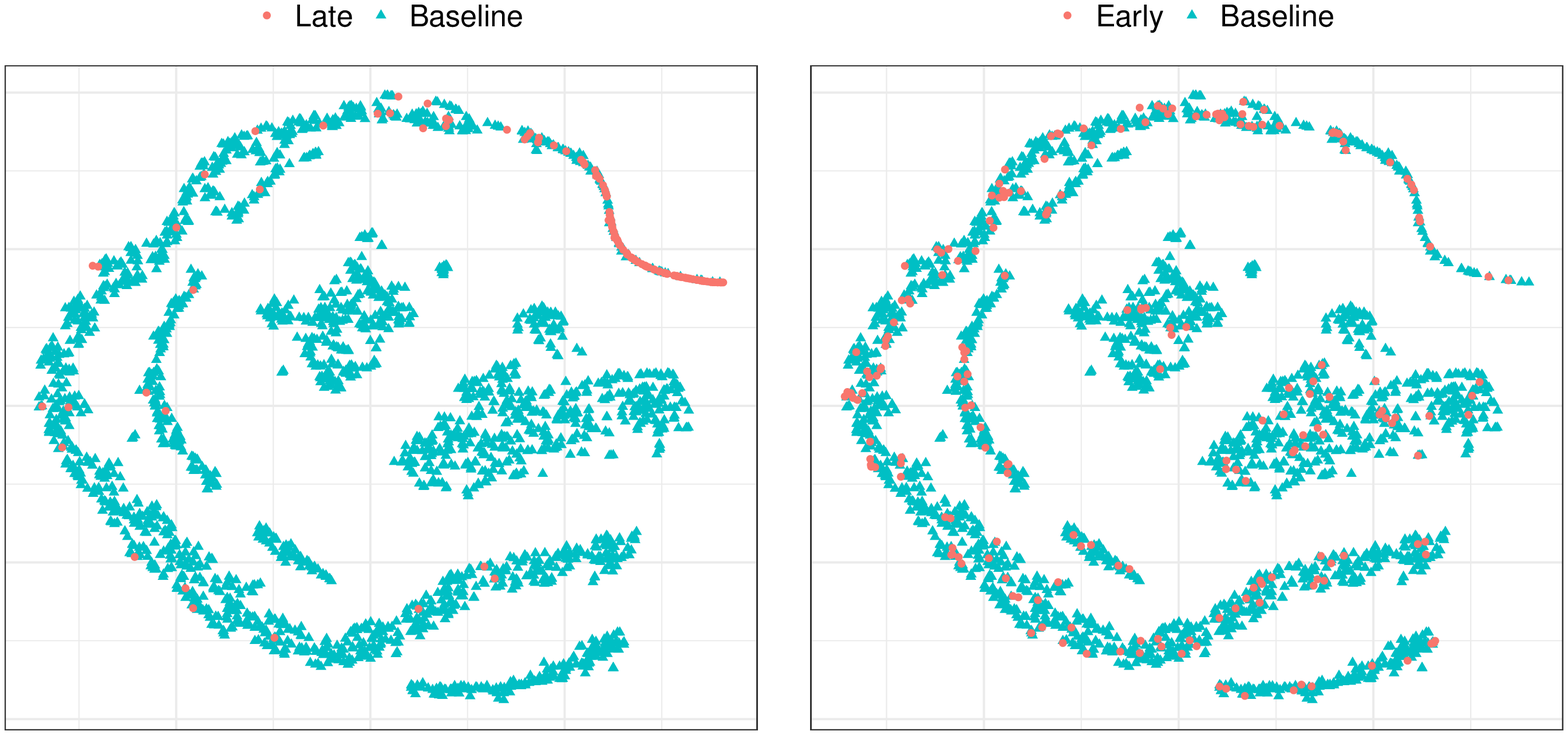}
\caption{\small A t-SNE plot of the data for Day 1. The $d = 16$ playing attributes are projected to a two dimensional space.}
\label{fig:realdata_3}
%\end{figure}
%\begin{figure}[!h]
\centering
\includegraphics[width=0.8\linewidth]{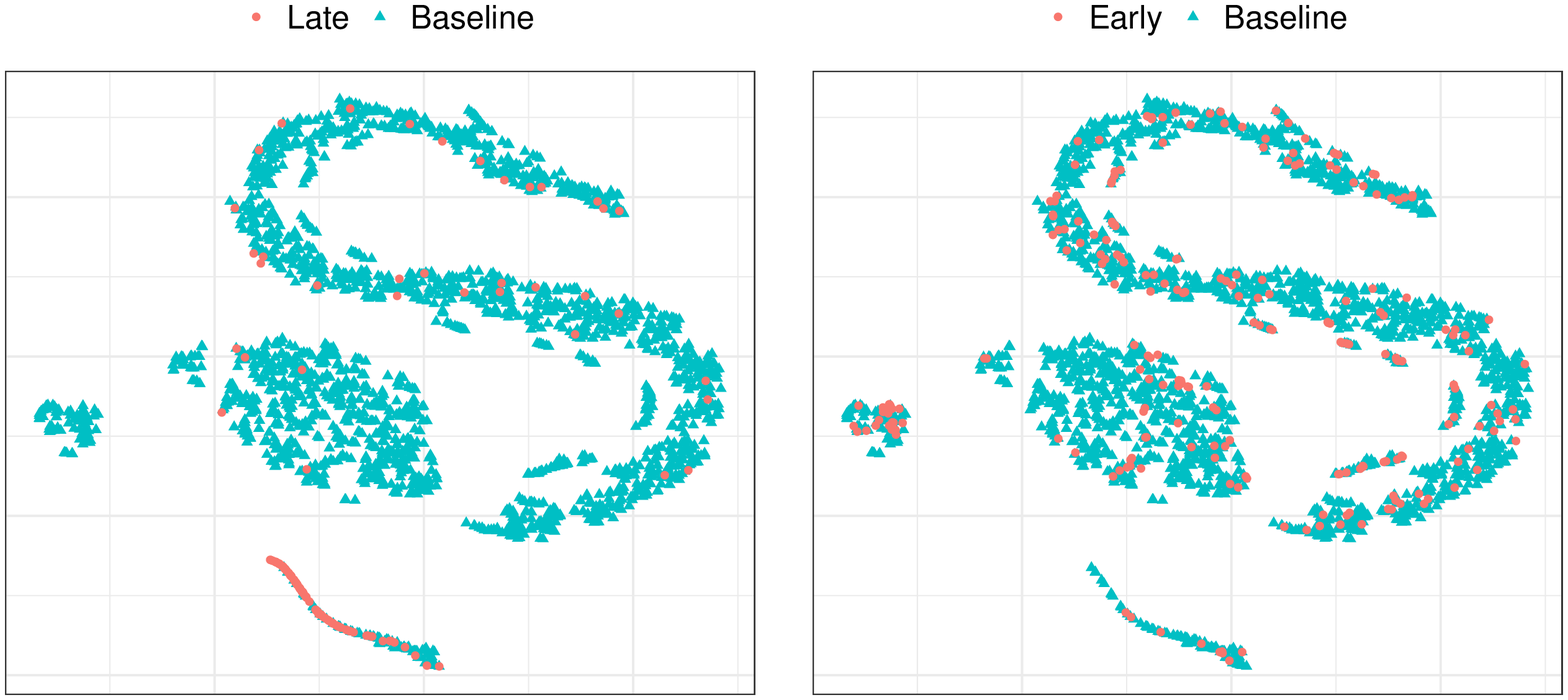}
\caption{\small A t-SNE plot of the data for Day 4. The $d = 16$ playing attributes are projected to a two dimensional space.}
\label{fig:realdata_1}
\end{figure}

Figure \ref{fig:realdata_boxplot} provides a box-plot of the 16 playing characteristics on day 5. It reveals that \texttt{Late} players seem to play relatively larger number of games (num\_games), spend more time playing with the game environment (pve\_quests and pve\_time) and are relatively more engaged with their friends and social connections within the game (numfrnd-games,numfrnd-played, {numguild-played}) than their counterparts in \texttt{Baseline}. The \texttt{Early} players, on the other hand, do not exhibit such stark contrasts in their playing behavior when compared to the \texttt{Baseline}. A t-SNE plot \citep{van2008visualizing} of the $d=16$ dimensional data in figures \ref{fig:realdata_3} and \ref{fig:realdata_1} provide further insights into the behavior of the \texttt{Late} and \texttt{Early} players. For both days 1 and 4, these figures exhibit the underlying heterogeneity in the \texttt{Baseline} player sample. Moreover, the \texttt{Late} sample occupies a distinct position in the two dimensional space that is away from the bulk of the \texttt{Baseline} sample. 

Tables \ref{tab:addicted_ pvals} and \ref{tab:engaged_pvals} report the $p-$values for the two testing problems. In Table \ref{tab:addicted_ pvals}, all competing tests, with the exception of TRUH, reject the null hypothesis $H_{01}: F^{(1)}_{\bm Y_1}\in\mathcal{F}(F_{\bm X})$ across the $7$ days and conclude that the playing behavior of the \texttt{Late} players is significantly different from \texttt{Baseline}. This corroborates the visual evidence found in figures \ref{fig:realdata_boxplot}, \ref{fig:realdata_3} and \ref{fig:realdata_1} that indeed the players who login late differ from their counterparts in \texttt{Baseline} as far as these $16$ playing characteristics are concerned. 
\begin{table}[!h]
\renewcommand{\arraystretch}{0.6}
\centering
\caption{\small $p-$values for testing $H_{01}: F^{(1)}_{\bm Y_1}\in\mathcal{F}(F_{\bm X})~vs~H_{11}:  F^{(1)}_{\bm Y_1}\notin \mathcal{F}(F_{\bm X})$}
\scalebox{0.9}{
	\begin{tabular}{ccccccccc}
		\hline
		Day & $n$ &$m_1$  & \multicolumn{1}{c}{EC test} &\multicolumn{1}{c}{GEC test} & \multicolumn{1}{c}{WEC test} & \multicolumn{1}{c}{TRUH}&\multicolumn{1}{c}{BGEC test} & \multicolumn{1}{c}{BWEC test} \\
		\hline
		1  &$2,558$& $133$   & $<10^{-3}$  &  $<10^{-3}$   &  $<10^{-3}$  &  $1.000$ &   $<10^{-3}$    & $<10^{-3}$ \\
		2  &$2,374$& $103$  &$<10^{-3}$ &  $<10^{-3}$    &   $<10^{-3}$    &  $<10^{-3}$ & $<10^{-3}$  &$<10^{-3}$\\
		3  &$2,073$&  $163$ & $<10^{-3}$ &  $<10^{-3}$    &   $<10^{-3}$    &   $1.000$    & $<10^{-3}$ & $<10^{-3}$\\
		4  & $2,291$ & $126$ & $<10^{-3}$ &   $<10^{-3}$   & $<10^{-3}$      & $<10^{-3}$      & $<10^{-3}$ &$<10^{-3}$\\
		5   &$2,340$& $143$  &$<10^{-3}$ & $<10^{-3}$      &   $<10^{-3}$    &  $1.000$     & $<10^{-3}$ & $<10^{-3}$\\
		6 &$2,560$& $72$   & $<10^{-3}$ &   $<10^{-3}$   &   $<10^{-3}$     &  $1.000$     & $<10^{-3}$  & $<10^{-3}$\\
		7   & $2,268$& $140$  & $<10^{-3}$&  $<10^{-3}$    &   $<10^{-3}$   &   $<10^{-3}$    & $<10^{-3}$ & $<10^{-3}$\\
		\hline
\end{tabular}}%
\label{tab:addicted_ pvals}%
\end{table}%
\begin{table}[!h]
\renewcommand{\arraystretch}{0.6}
\centering
\caption{\small $p-$values for testing $H_{02}: F^{(2)}_{\bm Y_2}\in\mathcal{F}(F_{\bm X})~vs~H_{12}: F^{(2)}_{\bm Y_2}\notin \mathcal{F}(F_{\bm X})$}
\scalebox{0.9}{
	\begin{tabular}{ccccccccc}
		\hline
		Day  &$n$ &$m_2$ & \multicolumn{1}{c}{EC test} &\multicolumn{1}{c}{GEC test} & \multicolumn{1}{c}{WEC test} & \multicolumn{1}{c}{TRUH}&\multicolumn{1}{c}{BGEC test} & \multicolumn{1}{c}{BWEC test} \\
		\hline
		1   & $2,558$& $203$ & $0.702$ &  $<10^{-3}$    &  $<10^{-3}$   & $0.525$ & $0.863$     & $0.917$\\
		2 &$2,374$&  $223$ & $0.193$  &$<10^{-3}$  & $<10^{-3}$ &   $0.995$ & $1.000$  & $1.000$\\
		3  &$2,073$&$171$ & $0.864$ & $0.001$     &  $0.002$     &    $0.715$   & $1.000$ & $1.000$ \\
		4  &$2,291$& $232$ & $<10^{-3}$ & $<10^{-3}$      & $<10^{-3}$  &  $0.795$     & $1.000$ & $1.000$\\
		5  &$2,340$& $216$   & $<10^{-3}$ & $<10^{-3}$  & $<10^{-3}$ & $0.105$ & $0.995$  &$0.998$\\
		6 &$2,560$&$177$ &$0.487$ & $<10^{-3}$     &   $<10^{-3}$    &   $0.890$    & $1.000$  & $1.000$\\
		7   &$2,268$& $199$ & $0.708$ & $<10^{-3}$     &    $<10^{-3}$   &   $0.855$    & $0.978$ & $0.988$\\
		\hline
\end{tabular}}%
\label{tab:engaged_pvals}%
\end{table}%
TRUH, on the other hand, does not provide such a consistent picture across the seven days and fails to reject $H_{01}$ in four out of the seven days. The numerical experiments of Section \ref{sec:sims} reveal that TRUH is relatively less powerful than BGEC and BWEC tests in detecting departures from $H_{01}$ and the result in Table \ref{tab:addicted_ pvals} is potentially another empirical evidence in that direction. The $p-$values reported in Table \ref{tab:engaged_pvals} exhibit an interesting pattern for the testing problem $H_{02}: F^{(2)}_{\bm Y_2}\in\mathcal{F}(F_{\bm X})~vs~H_{12}: F^{(2)}_{\bm Y_2}\notin \mathcal{F}(F_{\bm X})$. We note that TRUH, BWEC and BGEC tests fail to reject $H_{02}$ across the seven days, while the decisions from GEC and WEC tests are exactly the opposite, potentially demonstrating the non-conservativeness of these tests for testing the composite null hypothesis $H_{02}$. On five out of the seven days, EC test fails to reject $H_{02}$ and gives the impression that it is conservative for testing $H_{02}$. However, and as observed in Section \ref{sec:sims}, at moderately high dimensions the EC test statistic suffers from variance boosting under sample size imbalance and demonstrates low power.
\section{Discussion}
\label{sec:discuss}
%Nonparametric two-sample testing is a popular statistical tool for detecting distributional differences between the two samples without imposing any parametric assumptions on the underlying distributions. However, the multivariate samples from
The multivariate samples from modern gargantuan datasets often involve heterogeneity and direct application of existing nonparametric two-sample tests, such as the edge count tests, may lead to incorrect scientific decisions if they are not properly calibrated for the underlying latent heterogeneity. In this article, we demonstrate that under a composite null hypothesis, that allows the possibility that two samples may originate from the same mixture distribution but with possibly different mixing weights, the weighted edge count test can be re-calibrated to obtain a test which is asymptotically powerful for general alternatives. For practical implementation of this test, we develop a bootstrap calibrated weighted edge count test and demonstrate its excellent finite sample properties across a wide range of simulation studies. On a real world video game dataset, we use our testing procedure to detect addictive playing behavior and find that in comparison to players who exhibit normal playing behavior, players who login late to the game exhibit aberrant behavior while players who login early do not, thus confirming some of the findings in existing literature related to video game addiction.

Our future research will be directed towards extending the proposed testing framework in two directions. First, it will be interesting to develop an extension of the kernel two-sample tests of \cite{gretton2007kernel} to the heterogeneous setting and devise an efficient re-calibration procedure for these tests for consistent two-sample testing under the composite null hypothesis of Equation \eqref{eq:setup-3}. Second, for dealing with samples that involve high dimensional features, such as single cell RNA-seq data where $d\sim 10^4$, or data on consumer preferences for high dimensional product attributes, we will be interested in developing an extension of our testing procedure that allows incorporating such data-types into our heterogeneous two-sample testing framework.

\section*{Acknowledgements}
T. Banerjee was partially supported by the University of Kansas General Research Fund allocation \#2302216. B. Bhattacharya was partially supported by NSF CAREER DMS 2046393 and NSF DMS 2113771.

\spacingset{0.6} 
\bibliographystyle{plainnat}
\bibliography{ref,single-cell,new}
\newpage
\spacingset{1.5} 
\begin{center}
	{\large\bf SUPPLEMENTARY MATERIAL FOR ``Bootstrapped Edge Count Tests for Nonparametric
		Two-Sample Inference Under Heterogeneity"}
\end{center}
\appendix
This supplement is organized as follows: Section \ref{sec:pfconsistency} provides the proof of Proposition \ref{consistency}, Section \ref{app:figure_1_2_details} includes the simulation scheme for figures \ref{fig:motfig_1} and \ref{fig:motfig_2}, Section \ref{app:exp3} presents an additional simulation setting for the comparing the performance of \texttt{BWEC}, \texttt{BGEC}, and Section \ref{app:data} provides the data dictionary for the data used in Section \ref{sec:realdata}.
\section{Proof of Proposition~\ref{consistency}}
\label{sec:pfconsistency}
Recall from Equation \eqref{eq:deltaxy} the definition of the Henze-Penrose divergence: 
\begin{align*}%\label{eq:deltaxypf}
	\delta_\rho(f_{\bm X}, f_{\bm Y}) = \int_S \frac{ \rho f_{\bm X}^2(x) + f_{\bm Y}^2(x)}{( \rho f_{\bm X}(x) + f_{\bm Y}(x))}\mathrm dx .
\end{align*} 
%(Note that $\delta(f, g,\rho) \geq \delta(f, f,\rho) = \frac{1+\rho^2}{1+\rho}$.)
We begin with the following lemma: 
\begin{lemma}\label{lm:fxy} Under the assumptions of Proposition \ref{consistency}, for any $f_{\bm Y} \in \mathcal F(f_{\bm X})$, 
	$$\delta_\rho(f_{\bm X}, f_{\bm Y}) <  1 + \frac{(1+\rho)K^2}{L \rho^2} . $$ 
\end{lemma}

\textit{Proof:} Define the function $r_\rho: \R_{\geq 0}^2 \rightarrow \R_{\geq 0}$ as:   
$$r_\rho(s, t) =  \frac{\rho s^2 + t^2}{(\rho s + t)}.$$ 
Note that $\delta_\rho(f_{\bm X}, f_{\bm Y}) = \int_S r_\rho(f_{\bm X}(x), f_{\bm Y}(x)) \mathrm d x$. By a Taylor series expansion of the function 
$t \rightarrow r_\rho(f_{\bm X}(x), t) $ we can write, 
\begin{align}\label{eq:derivativepf}
	& r_\rho(f_{\bm X}(x), f_{\bm Y}(x)) - r_\rho(f_{\bm X}(x), f_{\bm X}(x)) \nonumber \\ 
	& = \frac{\partial}{\partial t} r_\rho(f_{\bm X}(x), t)\mid_{t=f_{\bm X}(x)} (f_{\bm Y}(x)-f_{\bm X}(x)) + \frac{(f_{\bm Y}(x)-f_{\bm X}(x))^2}{2}   \frac{\partial^2}{\partial t^2} r_\rho(f_{\bm X}(x), t)\mid_{t=\zeta_x} , 
\end{align} 
for some $\zeta_x \in [f_{\bm X}(x) \wedge f_{\bm Y}(x), f_{\bm X}(x) \vee f_{\bm Y}(x)]$. Note that 
$$\frac{\partial}{\partial t} r_\rho(f_{\bm X}(x), t) =  \frac{t^2 + 2 \rho  f_{\bm X}(x) t - \rho f_{\bm X}(x)^2}{(\rho f_{\bm X}(x) +  t)^2}  \quad \text{and} \quad \frac{\partial^2}{\partial^2 t} r_\rho(s, t) =  \frac{2 \rho (1+\rho) f_{\bm X}(x)^2}{(\rho f_{\bm X}(x) + t)^3}.$$ 
This implies, $\frac{\partial}{\partial t} r_\rho(f_{\bm X}(x), t)\mid_{t=f_{\bm X}(x)} = \frac{1}{1+\rho}:=c_\rho$. Then, using  $r_\rho(f_{\bm X}(x), f_{\bm X}(x)) = f_{\bm X}(x)$ Equation \eqref{eq:derivativepf} simplifies to 
\begin{align*}
	r_\rho(f_{\bm X}(x), f_{\bm Y}(x)) 
	%& = \frac{\partial}{\partial t} r_\rho(f_{\bm X}(x), t)\mid_{t=f_{\bm X}(x)} (f_{\bm Y}(x)-f_{\bm X}(x)) + \frac{1}{2} (f_{\bm Y}(x)-f_{\bm X}(x))^2  \frac{\partial^2}{\partial t^2} r_\rho(f_{\bm X}(x), t)\mid_{t=\zeta_x} \nonumber \\ 
	& =  f_{\bm X}(x) + c_\rho (f_{\bm Y}(x)-f_{\bm X}(x))  + (f_{\bm Y}(x)-f_{\bm X}(x))^2 \frac{\rho (1+\rho) f_{\bm X}^2(x)}{(\rho f_{\bm X}(x)+ \zeta_x)^3} . 
\end{align*}
Integrating both sides over $x \in S$ and using $\int_S f_{\bm X} \mathrm d x =\int_S f_{\bm Y}(x) \mathrm d x =1$, gives
\begin{align} 
	\delta_\rho(f_{\bm X}, f_{\bm Y}) & = 1 + \rho (1+\rho) \int_S (f_{\bm Y}(x)-f_{\bm X}(x))^2 \frac{ f_{\bm X}^2(x)}{ (\rho f_{\bm X}(x)+ \zeta_x)^3} \mathrm dx \label{eq:fderivativeintegration} \\ 
	& \leq 1+ \frac{1+\rho}{\rho^2} \int_S \frac{(f_{\bm Y}(x)-f_{\bm X}(x))^2}{f_{\bm X}(x)} \mathrm dx , \label{eq:derivativeintegration}
\end{align}
where the last step uses $\rho f_{\bm X}(x)+ \zeta_x \geq \rho f_{\bm X}(x)$. Now, suppose $f_{\bm Y} \in \mathcal F(f_{\bm X})$, that is, $f_{\bm Y} = \sum_{a=1}^K \lambda_a f_a(x)$, for some $\lambda_1, \lambda_2, \ldots, \lambda_K \in [0, 1]$ such that $\sum_{a=1}^K \lambda_a = 1$. Also, recall from Equation \eqref{eq:setup-1} that $f_{\bm X} = \sum_{a=1}^K w_a f_a(x)$. Then for $f_{\bm Y} \in \mathcal F(f_{\bm X})$ we have from Equation \eqref{eq:derivativeintegration}, 
\begin{align}\label{eq:integrationpf}
	\delta_\rho(f_{\bm X}, f_{\bm Y}) & = 1+ \frac{1+\rho}{\rho^2} \int_S \frac{(\sum_{a=1}^K (\lambda_a - w_a) f_a(x) )^2 }{\sum_{a=1}^K w_a f_a(x)} \mathrm dx \nonumber \\ 
	& \leq 1 + \frac{(1+\rho)K}{\rho^2}  \int_S \frac{\sum_{a=1}^K (\lambda_a - w_a)^2 f_a^2(x) }{\sum_{a=1}^K w_a f_a(x)} \mathrm dx , 
\end{align}
where the last step follows from the Cauchy-Schwarz inequality. Note that by Assumption \ref{assumption:w}, 
$$\sum_{a=1}^K w_a f_a(x)  > L \sum_{a=1}^K f_a(x) .$$
Moreover, 
$$\sum_{a=1}^K (\lambda_a - w_a)^2 f_a^2(x) \leq \sum_{a=1}^K f_a^2(x) \leq 
\left(\sum_{a=1}^K f_a(x)\right)^2.$$ Using the above two bounds in Equation \eqref{eq:integrationpf} we have, 
\begin{align*}
	\delta_\rho(f_{\bm X}, f_{\bm Y}) & < 1 + \frac{(1+\rho)K}{L \rho^2} \int_S  \sum_{a=1}^K f_a(x) \mathrm dx = 1 + \frac{(1+\rho)K^2}{L \rho^2} . 
\end{align*} 
This completes the proof of Lemma \ref{lm:fxy}. 

To complete the proof of the first statement in Proposition \ref{consistency} recall from Equation \eqref{eq:Rxynm} that for $f_{\bm Y} \in \mathcal F(f_{\bm X})$, 
\begin{equation}\label{eq:Rfxfy} 
	\mathcal{R}_w(\mathcal X_n,\mathcal Y_m) \pto \frac{\ell  \rho}{ (1+\rho)^2} \cdot \delta_\rho(f_{\bm X}, f_{\bm Y}) <  \frac{\ell  \rho}{ (1+\rho)^2} \left( 1 + \frac{(1+\rho)K^2}{L \rho^2} \right) := \gamma.
\end{equation} 
This implies, 
$$\lim_{m, n \rightarrow \infty}\P_{f_{\bm X}, f_{\bm Y}}(\mathcal{R}_w(\mathcal X_n,\mathcal Y_m) >  \gamma) = 0 ,$$ 
for $f_{\bm Y} \in \mathcal F(f_{\bm X})$. 
%Therefore, since $\hat \gamma_{m, n}/\gamma \stackrel{P} \rightarrow  1$, $$\lim_{m, n \rightarrow \infty}\P_{f_{\bm X}, f_{\bm Y}}(T_{m, n} > (1+\varepsilon) \hat \gamma_{m, n}) = 0.$$ 

For proving the second statement in Proposition \ref{consistency}, using $\zeta_x \in [f_{\bm X}(x) \wedge f_{\bm Y}(x), f_{\bm X}(x) \vee f_{\bm Y}(x)]$ and Assumption \ref{assumption:fxy} in Equation \eqref{eq:fderivativeintegration} we have, 
\begin{align}
	\delta_\rho(f_{\bm X}, f_{\bm Y}) & = 1 + \rho (1+\rho) \int_S (f_{\bm Y}(x)-f_{\bm X}(x))^2 \frac{ f_{\bm X}^2(x)}{ (\rho f_{\bm X}(x)+ \zeta_x)^3} \mathrm dx \nonumber \\ 
	& \geq 1 + \rho (1+\rho) \int_S (f_{\bm Y}(x)-f_{\bm X}(x))^2 \frac{ f_{\bm X}^2(x)}{( \rho f_{\bm X}(x)+f_{\bm X}(x) \vee f_{\bm Y}(x) )^3} \mathrm dx \nonumber \\ 
	\label{eq:M2}  & \geq 1 + \frac{\rho }{M_2^3 (1+\rho)^2} \int_S (f_{\bm Y}(x)-f_{\bm X}(x))^2  f_{\bm X}^2(x) \mathrm dx \\ 
	%& \geq1 + \rho (1+\rho) M_1^2 \int_S (f_{\bm Y}(x)-f_{\bm X}(x))^2 \frac{ 1 }{(f_{\bm X}(x)+ (\rho f_{\bm X}(x) \vee f_{\bm Y}(x) )^3} \mathrm dx \nonumber \\ 
	\label{eq:M1}  & \geq 1 + \frac{\rho M_1^2}{M_2^3 (1+\rho)^2} \int_S (f_{\bm Y}(x)-f_{\bm X}(x))^2  \mathrm dx . 
	%& \gtrsim \frac{2 \rho}{(1+\rho)} \int_S (f_{\bm Y}(x)-f_{\bm X}(x))^2 \frac{ 1 }{ f_{\bm X}(x) + f_{\bm X}(x) \vee \frac{g^3(x)}{f_{\bm X}^2(x)}} \mathrm dx \nonumber \\ 
	%& \geq \frac{\rho}{4 M(1+\rho) } \int_S (f_{\bm Y}(x)-f_{\bm X}(x))^2   \frac{ 1 }{ 1+ 1 \vee \frac{g^2(x)}{f_{\bm X}^2(x)}} \mathrm dx . 
\end{align} 
Note that the bound in Equation \eqref{eq:M2} uses $\sup_{x \in S} \left\{ \rho f_{\bm X}(x) +  f_{\bm X}(x) \vee f_{\bm Y}(x)) \right\} \leq M_2 (1+\rho)$ and the bound in Equation \eqref{eq:M1} uses $f_{\bm X}(x) \geq M_1$, for all $x \in S$, by Assumption \ref{assumption:fxy}. 

Now, if $f_{\bm Y} \notin \mathcal F(f_{\bm X})$ is such that  $\int_S (f_{\bm Y}(x)-f_{\bm X}(x))^2  \mathrm dx \geq \varepsilon^2:= \frac{2 M_2^3 (1+\rho)^3 K^2}{ M_1^2 \rho^3 L}$, then 
\begin{equation*}%\label{eq:Rxynm} 
	\mathcal{R}_w(\mathcal X_n,\mathcal Y_m) \pto \frac{\ell  \rho}{ (1+\rho)^2} \cdot \delta_\rho(f_{\bm X}, f_{\bm Y}) \geq \gamma + \frac{\ell K^2}{L \rho (1 + \rho)} , 
\end{equation*} 
for $\gamma$ as defined in Equation \eqref{eq:Rfxfy}. Therefore, 
\begin{equation}\label{eq:Rxymn}
	\lim_{m, n \rightarrow \infty}\P_{f_{\bm X}, f_{\bm Y}}(\mathcal{R}_w(\mathcal X_n,\mathcal Y_m) > \gamma) = 1, 
\end{equation} 
for $f_{\bm Y} \notin \mathcal F(f_{\bm X})$ such that $\int_S (f_{\bm Y}(x)-f_{\bm X}(x))^2  \mathrm dx \geq \varepsilon^2$. 
\hfill $\Box$ 

% 
%\begin{remark} {\em Define the function $h(t)=\frac{(x-1)^2}{(x+1)^3}$. Then 
%$$\int_S (f_{\bm Y}(x)-f_{\bm X}(x))^2 \frac{ f_{\bm X}^2(x)}{(f_{\bm X}(x) + f_{\bm Y}(x) )^3} \mathrm dx  = \int_S  h\left(\frac{f_{\bm Y}(x)}{f_{\bm X}(x)}\right) f_{\bm X}(x) \mathrm d x.$$ }
%\end{remark} 
% 

\begin{remark}\label{remark:assumption} In the proof of Proposition \ref{consistency}  the assumption that $f_{\bm X}$ is bounded away from zero is used in Equation \eqref{eq:M1}. This assumption can be removed if instead we assume that there exists $\kappa > 0$ such that the truncated $L_2$-distance: 
\begin{align}\label{eq:assumptiontruncation}
	\int_{ \{ f_{\bm X}(x) > \kappa \} } (f_{\bm Y}(x)-f_{\bm X}(x))^2 \geq \tilde \varepsilon^2/\kappa^2 , 
\end{align} 
where $\tilde \varepsilon^2:= \frac{2 M_2^3 (1+\rho)^3 K^2}{ \rho^3 L}$. To see this, note from Equation \eqref{eq:M2} that for any $\kappa > 0$, 
\begin{align}
	\delta_\rho(f_{\bm X}, f_{\bm Y}) & \geq 1 + \frac{\rho }{M_2^3 (1+\rho)^2} \int_S (f_{\bm Y}(x)-f_{\bm X}(x))^2  f_{\bm X}^2(x) \mathrm dx \nonumber \\ 
	& \geq 1 + \frac{\rho \kappa^2}{M_2^3 (1+\rho)^2} \int_{\{ f_{\bm X}(x) > \kappa \} } (f_{\bm Y}(x)-f_{\bm X}(x))^2  \mathrm dx \geq 1+ \frac{2(1+\rho)K^2}{L \rho^2} , \nonumber 
\end{align} 
where the last inequality uses Equation \eqref{eq:assumptiontruncation} and the definition of $\varepsilon$. This implies, 
\begin{equation*}%\label{eq:Rxynm} 
	\mathcal{R}_w(\mathcal X_n,\mathcal Y_m) \pto \frac{\ell  \rho}{ (1+\rho)^2} \cdot \delta_\rho(f_{\bm X}, f_{\bm Y}) \geq \gamma + \frac{\ell K^2}{L \rho (1 + \rho)} . 
\end{equation*} 
Therefore, Equation \eqref{eq:Rxymn} holds whenever the truncated $L_2$-condition in Equation \eqref{eq:assumptiontruncation} is satisfied.
\end{remark}
\section{Details for figures \ref{fig:motfig_1} and \ref{fig:motfig_2}}
\label{app:figure_1_2_details}
We next describe the two simulation settings that were discussed in Section \ref{sec:motivation} under examples 1, 2 and figures \ref{fig:motfig_1}, \ref{fig:motfig_2}. For Example 1 and Figure \ref{fig:motfig_1}, we fix $n=2000,~m=200,~d=2$ and let $F_{\bm X}=\sum_{k=1}^{4}w_k\mathcal{N}_d(\bm \mu_k,\bm I_d)$ where $w_k=0.25$ for $k=1,\ldots,4$, $\bm \mu_1=(10,10)$, $\bm \mu_2=(20,10)$, $\bm \mu_3=(20,20)$ and $\bm \mu_4=(10,20)$. We consider three scenarios for simulating $\mathcal Y_m$ from $F_{\bm Y}$. For Case I, we let $F_{\bm Y}=F_{\bm X}$ and ,thus, $F_{\bm Y}\in\mathcal{F}(F_{\bm X})$ since $F_{\bm Y}$ has all the subpopulations present in $F_{\bm X}$. In Case II, we consider $F_{\bm Y}=0.1\,\mathcal{N}_d(\bm\mu_1,\bm I_d)+0.8\,\mathcal{N}_d(\bm\mu_2,\bm I_d)+0.1\,\mathcal{N}_d(\bm\mu_3,\bm I_d)$. So, $F_{\bm Y}\in\mathcal{F}(F_{\bm X})$ since $F_{\bm Y}$ has all the subpopulations present in $F_{\bm X}$ but at different proportions. Finally, for Case III we let  $F_{\bm Y}=\mathcal{N}_d(\bm\mu_5,\bm I_d)$ where $\bm \mu_5=(25,5)$. This setting presents a scenario where $F_{\bm Y}\notin\mathcal{F}(F_{\bm X})$ and the composite null $H_0$ is not true. 
Table \ref{tab:motfig_1} reports the rejection rates for testing $H_0:F_{\bm Y}\in\mathcal{F}(F_{\bm X})~vs~H_1:F_{\bm Y}\notin\mathcal{F}(F_{\bm X})$ under the three scenarios described in Example 1 and Figure \ref{fig:motfig_1}. We note that the three edge count tests cannot distinguish Case II from Case III and infer $F_{\bm Y}\notin\mathcal{F}(F_{\bm X})$ for both these cases.%, thus suggesting their inability to tackle subpopulation level heterogeneity.
\begin{table}[!h]
\renewcommand{\arraystretch}{0.7}
\centering
\caption{{\small Rejection rates at $5\%$ level of significance: Example 1 and Figure \ref{fig:motfig_1} in Section \ref{sec:motivation}}.}
\scalebox{0.9}{
	\begin{tabular}{lccc}
		\hline
		& \multicolumn{3}{c}{$n = 2000,~m = 200,~d=2$} \\
		\hline
		\multicolumn{1}{c}{Method} & Left panel & Center panel& Right panel\\
		\multicolumn{1}{c}{} & Case I - $F_{\bm Y}\in\mathcal{F}(F_{\bm X})$   & Case II - $F_{\bm Y}\in\mathcal{F}(F_{\bm X})$    & Case III - $F_{\bm Y}\notin\mathcal{F}(F_{\bm X})$    \\
		\hline
		\texttt{EC} test & 0.048 & 1.000 & 1.000 \\
		\texttt{GEC} test & 0.032 &  1.000 & 1.000 \\
		\texttt{WEC} test & 0.056 &  1.000 & 1.000 \\
		\hline
		\texttt{TRUH} test  & 0.048 & 0.058 & 1.000 \\
		\hline
		\texttt{BGEC} test & 0.000 & 0.000 & 1.000 \\
		\texttt{BWEC} test & 0.000 & 0.000 & 1.000  \\
		\hline
\end{tabular}}%
\label{tab:motfig_1}%
\end{table}%

For Example 2 and Figure \ref{fig:motfig_2}, we take $d=3$ and let $F_{\bm X}=\sum_{k=1}^{3}w_k\mathcal{N}_d(\bm \mu_k,\bm I_d)$ where $w_1=w_2=0.3,~w_3=0.4$, $\bm \mu_1=(0,0,0)$, $\bm \mu_2=(0,-4,-4)$, and $\bm \mu_3=(4,-2,-3)$. We consider three scenarios for simulating $\mathcal Y_m$ from $F_{\bm Y}$. In Case I, we let $F_{\bm Y}=F_{\bm X}$ and, thus, $F_{\bm Y}\in\mathcal{F}(F_{\bm X})$. For Case II, we consider $F_{\bm Y}=0.8\,\mathcal{N}_d(\bm\mu_1,\bm I_d)+0.1\,\mathcal{N}_d(\bm\mu_2,\bm I_d)+0.1\,\mathcal{N}_d(\bm\mu_3,\bm I_d)$. So, $F_{\bm Y}\in\mathcal{F}(F_{\bm X})$ since $F_{\bm Y}$ has all the subpopulations present in $F_{\bm X}$ but at different proportions. In Case III, we let  $F_{\bm Y}=0.8\,\mathcal{N}_d(\bm\mu_1,\bm \Sigma_d)+0.1\,\mathcal{N}_d(\bm\mu_2,\bm \Sigma_d)+0.1\,\mathcal{N}_d(\bm\mu_3,\bm \Sigma_d)$ where $\bm \Sigma_d=0.1\bm I_d$. This setting presents a scenario where $F_{\bm Y}\notin\mathcal{F}(F_{\bm X})$ since the components of $F_{\bm Y}$ differ from the components of $F_{\bm X}$ with respect to their scale parameters. 
Table \ref{tab:motfig_2} reports the rejection rates for testing $H_0:F_{\bm Y}\in\mathcal{F}(F_{\bm X})~vs~H_1:F_{\bm Y}\notin\mathcal{F}(F_{\bm X})$ under the three scenarios described in Example 2 and Figure \ref{fig:motfig_2}. We continue to note that the three edge count tests cannot distinguish Case II from Case III and infer $F_{\bm Y}\notin\mathcal{F}(F_{\bm X})$ for both these cases. This suggests that the edge count tests are unable to tackle subpopulation level heterogeneity. Additionally, the \texttt{TRUH} test incorrectly infers $F_{\bm Y}\in\mathcal{F}(F_{\bm X})$ for Case III, thus demonstrating low power for detecting departures from $H_0$ when the components of $F_{\bm Y}$ and $F_{\bm X}$ differ only with respect to their scales.
\begin{table}[!h]
\renewcommand{\arraystretch}{0.7}
\centering
\caption{{\small Rejection rates at $5\%$ level of significance: Example 2 and Figure \ref{fig:motfig_2} in Section \ref{sec:motivation}}.}
\scalebox{0.9}{
	\begin{tabular}{lccc}
		\hline
		& \multicolumn{3}{c}{$n = 2000,~m = 200,~d=3$} \\
		\hline
		\multicolumn{1}{c}{Method} & Left panel & Center panel& Right panel\\
		\multicolumn{1}{c}{} & Case I - $F_{\bm Y}\in\mathcal{F}(F_{\bm X})$   & Case II - $F_{\bm Y}\in\mathcal{F}(F_{\bm X})$    & Case III - $F_{\bm Y}\notin\mathcal{F}(F_{\bm X})$    \\
		\hline
		\texttt{EC} test & 0.068 & 1.000 & 1.000 \\
		\texttt{GEC} test & 0.068 &  1.000 & 1.000 \\
		\texttt{WEC} test & 0.064 &  1.000 & 1.000 \\
		\hline
		\texttt{TRUH} test  & 0.012 & 0.018 & 0.004 \\
		\hline
		\texttt{BGEC} test & 0.000 & 0.000 & 1.000 \\
		\texttt{BWEC} test & 0.000 & 0.000 & 1.000  \\
		\hline
\end{tabular}}%
\label{tab:motfig_2}%
\end{table}%
\section{Numerical experiment 3}
\label{app:exp3}
We consider a setting where the samples $\mathcal X_n$ and $\mathcal Y_m$ exhibit zero inflation across the $d$ dimensions. 
Denote $\bm \delta_{\{\bm 0\}}=(\delta_{1\{0\}},\ldots,\delta_{d\{0\}})$ denote the $d-$dimensional vector of point masses at $0$. We let $F_{\bm X}= \bm p\delta_{\{\bm 0\}}+(\bm 1_d-\bm p)~\Bigl\{0.5~F_1+0.5~F_2\Bigr\},$ where $F_1={\mathrm{Gam}}_d(\text{shape}=5\bm 1_d,\text{rate}=\bm 1_d,\bm \Sigma_1)$, $F_2={\mathrm{Exp}}_d(\text{rate}=\bm 1_d,\bm \Sigma_2)$ and $\bm p=(p_1,\ldots,p_d)$ is the vector of probabilities that regulate the differential zero inflation across the $d$ dimensions. We sample the first $0.8d$ coordinates of $\bm p$ independently from $\text{Unif}(0.5, 0.6)$, and the remaining $0.2d$ coordinates are set to $0$. So the first $0.8d$ coordinates of $F_{\bm X}$ encounter zero inflation. Finally, $\bm \Sigma_1, \bm \Sigma_2$ are as described in Experiment 2 (Section \ref{sec:simexp2}) and $\mathcal{X}_n$ are sampled from $F_{\bm X}$ using the R-package \texttt{lcmix}. For simulating $\mathcal Y_m$ from $F_{\bm Y}$, we consider the following two scenarios:
\begin{itemize}
\item Scenario I: $F_{\bm Y}=\bm p\bm\delta_{\{\bm 0\}}+(\bm 1_d-\bm p)~\{0.2~F_1+0.8~F_2\}$ and so $H_0:F_{\bm Y}\in\mathcal{F}(F_{\bm X})$ is true.
\item Scenario II: $F_{\bm Y}=\bm q\delta_{\{\bm 0\}}+(\bm 1_d-\bm q)~\{0.5~{\mathrm{Gam}}_d(\text{shape}=5\bm 1_d,\text{rate}=1.5\bm 1_d,\bm \Sigma_1)+0.5~F_2\}$, where the first $0.8d$ coordinates of $\bm q$ are set to $0.3$ and the remaining $0.2d$ coordinates to $0$. Apart from the differential zero inflation between $F_{\bm Y}$ and $F_{\bm X}$, the rate parameter of the first component of $F_{\bm Y}$ is different from that of $F_1$. So in this scenario, $G\notin\mathcal{F}(F_0)$ and $H_0$ is false. 
\end{itemize}
\begin{table}[!h]
\renewcommand{\arraystretch}{0.6}
\centering
\caption{\small Rejection rates at $5\%$ level of significance: Experiment 3 and Scenario 1 wherein $H_0: F_{\bm Y} \in \mathcal{F}(F_{\bm X})$ is true.}
\begin{tabular}{lcccccc}
	\hline
	& \multicolumn{3}{c}{$n = 500,~m = 50$} & \multicolumn{3}{c}{$n = 2000,~m = 200$} \\
	\hline
	\multicolumn{1}{c}{Method} & $d=5$   & $d=15$  & $d=30$  & $d=5$   & $d=15$  & $d=30$ \\
	\hline
	\texttt{EC} test & 0.070 & 0.074 & {0.024} & 0.302 & 0.140 & 0.102 \\
	\texttt{GEC} test & 0.264 & 0.284 & 0.256 & 0.624 & 0.680 & 0.752 \\
	\texttt{WEC} test & 0.352 & 0.408 & 0.392 & 0.790 & 0.830 & 0.880 \\
	\hline
	\texttt{TRUH} test & 0.002 & 0.000 & 0.000 & 0.000 & 0.000 & 0.000 \\
	\hline
	\texttt{BGEC} test & 0.046 & 0.000 & 0.000 & 0.000 & 0.000 & 0.000 \\
	\texttt{BWEC} test & {0.058} & 0.000 & 0.000 & 0.000 & 0.000 & 0.000 \\
	\hline
\end{tabular}%
\label{tab:exp3set1}%
\end{table}%
Table \ref{tab:exp3set1} reports the rejection rates for the competing tests under Scenario I and reveals that under the setting of zero inflation, the WEC and GEC tests are not conservative. \texttt{TRUH}, \texttt{BWEC} and \texttt{BGEC} tests, on the other hand, report rejection rates that are either at or below the prespecified $0.05$ level, thus demonstrating their conservatism in testing $H_0: F_{\bm Y} \in \mathcal{F}(F_{\bm X})~vs~H_1:F_{\bm Y} \notin \mathcal{F}(F_{\bm X})$.
\begin{table}[!h]
\renewcommand{\arraystretch}{0.6}
\centering
\caption{\small Rejection rates at $5\%$ level of significance: Experiment 3 and Scenario 2 wherein $H_0: F_{\bm Y} \in \mathcal{F}(F_{\bm X})$ is false.}
\begin{tabular}{lcccccc}
	\hline
	& \multicolumn{3}{c}{$n = 500,~m = 10$} & \multicolumn{3}{c}{$n = 2000,~m = 40$} \\
	\hline
	\multicolumn{1}{c}{Method} & $d=5$   & $d=15$  & $d=30$  & $d=5$   & $d=15$  & $d=30$ \\
	\hline
	\texttt{EC} test& 0.106 & 0.000 & 0.000 & 0.420 & 0.000 & 0.000 \\
	\texttt{GEC} test & 0.250 &  0.604 & 0.878 & 0.732 & 0.986 & 1.000 \\
	\texttt{WEC} test & 0.344 &  0.526 & 0.704 & 0.828 & 0.978 & 1.000 \\
	\hline
	\texttt{TRUH} test & 0.012 & 0.004 & 0.012 & 0.008 & 0.000 & 0.000 \\
	\hline
	\texttt{BGEC} test & 0.124 & 0.390 & 0.772 & 0.198 & 0.874 & 0.998 \\
	\texttt{BWEC} test & 0.138 & 0.288 & 0.512 & 0.196 & 0.822 & 0.986 \\
	\hline
\end{tabular}%
\label{tab:exp3set2}%
\end{table}%
In Table \ref{tab:exp3set2} we report the rejection rates for Scenario II when the sample size imbalance is $0.02$ as opposed to $0.1$ in Scenario I. Under this challenging setting, we find that \texttt{BWEC} and \texttt{BGEC} tests are more powerful than \texttt{TRUH} and demonstrate competitive power to \texttt{WEC} and \texttt{GEC} tests when the sample sizes are relatively large. Table \ref{tab:exp3set1} gives the impression that at $d=30$, the \texttt{EC} test is conservative for testing $H_0$ under Scenario I. However, at such moderately high dimensions and under unequal sample sizes, the edgecount test statistic suffers from variance boosting and demonstrates low power, which explains its relatively low rejection rates in both tables \ref{tab:exp3set1} and \ref{tab:exp3set2}.
\section{Data dictionary}
\label{app:data}
Table \ref{app:tab_datadic} provides a description of the $d=16$ player characteristics available in the video game data used in Section \ref{sec:realdata}.
\begin{table}[!h]
\centering
\caption{Data dictionary}
\scalebox{0.8}{
	\begin{tabular}{cll}
		\hline
		Sl no. & \multicolumn{1}{c}{Variable name} & \multicolumn{1}{c}{Description} \\
		\hline
		1     & game\_level & player’s level in the game \\
		2     & pve\_quests & no. of quests a player accomplished in Player Versus Environment (PVE) mode \\
		3     & pve\_mission & no. of missions a player accomplished in PVE mode \\
		4     & pve\_time & player’s daily time spent in playing PVE mode in hours \\
		5     & num\_game & no. of PVE game rounds a player played in a day \\
		6     & purch\_count & no. of purchases a player made in a day \\
		7     & frnd\_count & no. of friends a player has \\
		8     & frnd\_level & mean level of a player’s friends \\
		9     & numfrnd\_purch & no. of times of a player’s friends made purchases \\
		10    & valfrnd\_purch & monetary value of all purchases made by a player’s friends \\
		11    & numfrnd\_played & no. of friends a player played with during game sessions \\
		12    & numfrnd\_games & no. of game sessions a player played with her friends \\
		13    & tenure & no. of days a player has been associated with the game \\
		14    & guild\_tenure & no. of days a player has been associated with a guild \\
		15    & numguild\_played & no. of guild members a player played with \\
		16    & numguild\_games & no. of game sessions a player played with guild members \\
		\hline
\end{tabular}}%
\label{app:tab_datadic}%
\end{table}%
\end{document}